\documentclass[11pt,twoside]{article}
\usepackage{PRIMEarxiv}
\raggedright

\usepackage{setspace}

\usepackage{natbib}
 \bibpunct[, ]{(}{)}{,}{a}{}{,}%

 \usepackage[title]{appendix}

\usepackage[utf8]{inputenc} 
\usepackage[T1]{fontenc}    
\usepackage{hyperref}       
\usepackage{url}            
\usepackage{booktabs}       
\usepackage{amssymb}
\usepackage{amsmath}
\usepackage{amsfonts}       
\usepackage{nicefrac}       
\usepackage{microtype}      
\usepackage{lipsum}
\usepackage{fancyhdr}       
\usepackage{graphicx}       
\graphicspath{{Images/}}

\usepackage{algorithm}
\usepackage{algorithmic}
\usepackage[svgnames,table,]{xcolor}
\usepackage{xltabular}
\usepackage{adjustbox}
\usepackage{subcaption}
\usepackage{caption}
\newcolumntype{C}{>{\centering\arraybackslash}X}
\newcolumntype{L}{>{\raggedright\arraybackslash}X}
\newcolumntype{Z}{>{\hsize=1\hsize\raggedright}X}
\newcolumntype{Y}{>{\centering\small\raggedright\arraybackslash}X}
\newcommand{\solutionset}{S}

\title{{\sc DiversiTree}: A New Method to Efficiently Compute Diverse Sets of Near-Optimal Solutions to Mixed-Integer Optimization Problems
}

\author{
  Izuwa Ahanor, Hugh Medal \\
  Department of Industrial and Systems Engineering, \\
  University of Tennessee-Knoxville\\
  \texttt{iahanor@vols.utk.edu, hmedal@utk.edu} \\
   \And
  Andrew C. Trapp \\
  Business School, Data Science Program \protect\\ Worcester Polytechnic Institute \\
  \texttt{atrapp@wpi.edu} \\
}

\begin{document}
\maketitle

\onehalfspacing

\begin{abstract}
While most methods for solving mixed-integer optimization problems compute a single optimal solution, a diverse set of near-optimal solutions can often lead to improved outcomes. We present a new method for finding a set of diverse solutions by emphasizing diversity within the search for near-optimal solutions. Specifically, within a branch-and-bound framework, we investigated parameterized node selection rules that explicitly consider diversity. Our results indicate that our approach significantly increases the diversity of the final solution set. When compared with two existing methods, our method runs with similar runtime as regular node selection methods and gives a diversity improvement between 12\% and 190\%.
In contrast, popular node selection rules, such as best-first search, in some instances performed worse than state-of-the-art methods by more than 35\% and gave an improvement of no more than 130\%. Further, we find that our method is most effective when diversity in node selection is continuously emphasized after reaching a minimal depth in the tree and when the solution set has grown sufficiently large. Our method can be easily incorporated into integer programming solvers and has the potential to significantly increase the diversity of solution sets.
\end{abstract}


\section{Introduction}
\label{sec:1introduction}

It is often important to find sets of near-optimal solutions to optimization problems rather than a single solution. In particular, for these multiple near-optimal solutions to be usable, they should be diverse, to ensure that decision makers are presented with a variety of options.

There has been increasing awareness of finding not only one but multiple optima due to increased computational capabilities in the last two decades \citep{bertsimas2016best}. Some specific applications in which it is desirable to find multiple near-optimal solutions to optimization problems include the correct identification of metabolic activity of cells and tissues in metabolic networks \citep{Rodriguez-Mier.2021}, aiding motif finding in computational molecular biology \citep{Zaslavsky.2006}, enabling exploration and mapping searches to broader but specific solutions in large search spaces by including near-optimal search results in search requests \citep{mouret2015illuminating, zahavy2021discovering}, and providing policies that are more robust to data changes in reinforcement learning and machine learning \citep{kumar2020one, eysenbach2018diversity, sharifnia2021robust}. Discovering multiple near-optimal solutions have also been applied to identifying alternative near-optimal structural designs \citep{He.2020}, adding more artistic alternatives to structural topology optimization \citep{Cai.2021, He.2020}, providing competitive alternatives to facility location and location routing problems \citep{Church.2020,  schittekat2009or}, diversifying software deployment to enable stronger computer software security \citep{Tsoupidi.2020}, generating multiple near-optimal group preferences in computational social choice analysis \citep{boehmer2021broadening}, and broadening architectural testing in processor design \citep{van2009constraint}.

Many of these applications have a large number of near-optimal solutions (see Table \ref{fig:problemsSolved2} of Appendix~\ref{Sec:trainingSet}). For instance, both the routing problem from \citet{ceria1998cutting} and the multi-period facility location problem studied in \citet{eckstein1996parallel} are known to have more than 10,000 solutions with objective value within 1\% of optimality. Some other problems like the image deblurring problem studied in \citet{biyouki2021blind} could have an infinite number of solutions. In problems with many near-optimal solutions, there is a need to identify a small subset of near-optimal solutions that is representative of the whole. One measure of how well a subset represents the whole is the \textit{diversity} of the subset. Unfortunately, methods for finding near-optimal solutions such as {\sc OneTree} find near-optimal sets consisting of solutions that are not very diverse (see example provided in \S \ref{sec:example}).

There are several important contexts in which it is useful to have a \textit{diverse} set of near-optimal solutions. First, in some design problems decision makers seek a set of designs from which to choose, for the purpose of considering other difficult-to-model factors when selecting a single design \citep{joseph2015sequential}. A similar context is in statistical model selection for a specific domain. Presenting a domain expert with a diverse set of models with similar fit allows the expert to select the model that best matches intuition. Several new studies hypothesize that for many machine learning tasks a set of models exist with near-minimal loss (\citet{semenova2019study}; also see \S 9 of \citet{rudin2022interpretable}). Second, using optimization for decision problems can often be an iterative process in which a MIP solution is first presented to a decision maker, only to have the decision maker identify that the solution violates an important side constraint that was not included in the model. If the decision maker is only provided a single solution, the model must be re-solved after adding the side constraint. However, if a diverse set of near-optimal solutions were readily available, the decision maker may be able to find a good solution in the set that does not violate the side constraint, avoiding the need to re-solve the model. Third, in contexts in which solutions are implemented repeatedly, it can be useful to alternate between a diverse set of near-optimal solutions. For example, one may use MIP to match workers to jobs to minimize total completion time, but regularly implement different near-optimal matchings to increase cross-training. Lastly, a set of near-optimal solutions can be used to measure and explain the importance of variables in applications such as statistical model selection (see \S 9 of \citet{rudin2022interpretable}). If a feature is present in a large number of diverse near-optimal solutions, this provides additional evidence that the feature is an important predictor of the response variable.

One class of methods for generating diverse sets of near-optimal solutions to mixed-integer optimization (MIO) problems uses a \textit{two-phase} approach (methods within this category typically also perform a precomputation to compute the optimal objective value. However, when referring to these methods as two-phase, we do not include the precomputation step as a phase). In the first \textit{solution generation} phase, an oracle finds a set of near-optimal solutions without considering diversity. For example, \cite{danna2007generating} developed the {\sc OneTree} oracle for this purpose. In the second \textit{diverse subset selection} phase the output set from the first phase is processed by heuristics (e.g., \cite{glover2000scatter,danna2009select}) or an optimization algorithm (e.g., \cite{danna2009select}) to select a small subset of the input set with maximum diversity. While this approach works well for problems with a small number of near-optimal solutions, for MIOs with a very large set of near-optimal solutions it is not practical to find the complete set. As a result, the first phase can only compute a subset of near-optimal solutions. Unfortunately, because existing first-phase methods do not consider diversity, the near-optimal sets they generate often lack diversity (see \S \ref{sec:4results}). If this subset is not diverse, then the smaller subset computed by the second phase necessarily will lack diversity. The work presented in this article addresses this issue.

\subsection{Related Work}
Although there exists some work on finding near-optimal solutions to continuous optimization problems (see \cite{lavine2019whim}), most of the work has been in relation to mixed-integer optimization (MIO) problems and, in particular, with respect to (binary) integer variables.

There are a number of different algorithms for generating a set of near-optimal solutions to a MIO problem. \cite{Achterberg.2008} developed an approach for generating all feasible solutions to an integer programming (IP) problem called \textit{branch-and-count}. This approach is based on detecting ``unrestricted subtrees" in the branching tree. This method can also be used to find all solutions within a certain threshold of the optimal objective value, if known. Other methods have been proposed for finding all optimal solutions; an example is the work by \cite{lee2000recursive} which uses a recursive MIP algorithm to find all alternate optima. \cite{Serra.2020} used weighted decision diagrams to compactly represent all near-optimal solutions generated for an integer programming problem. The compact representation eases resolves and information retrieval from all generated solutions. Also, there is the {\sc OneTree} method (currently implemented in GAMS-CPLEX) given by \cite{danna2007generating} which extends the branching tree used to solve the MIP problem to generate near-optimal solutions. In addition, approaches have been developed for specific classes of problems including using dynamic programming to generate multiple solutions to graph-based problems \citep{Baste.2019}, methods to represent near-optimal solutions compactly \citep{10.1007/s10107-019-01390-3} and algorithms specific to topology optimization problems \citep{wang2018diverse}. While these methods are effective at finding near-optimal solutions, they do not necessarily compute solution sets that are diverse. One of the few papers that discusses balancing diversity and optimality is \cite{Zhou.2016}. They developed a Dual Diverse Competitive Design (DDCD) method that formulates balancing optimality and diversity as an optimization problem that maximizes diversity, subject to constraints on the performance penalty. While their method specifically focused on generating competitive designs that would give diverse solutions, the goal of this current study is a method for general MIO problems.

There are a number of different approaches for finding a diverse set of near-optimal solutions to an optimization problem. In the \textit{sequential} approach \citep{danna2007generating}, the optimization problem is solved multiple times, and after each solve a constraint is added that requires the next optimal solution be different from the previous. The sequential approach is also used in \cite{trapp2015finding} to select diverse solutions to binary integer linear problems and in \cite{petit2015finding} in the context of constraint programming. Both methods consider maximizing a metric that is the ratio of diversity of solutions to loss in objective function. \cite{petit2019enriching} extended these ideas further by introducing the notion of infusing solutions with other desirable features such as fairness, persistence and balance when generating diverse near-optimal solutions. The \textit{variable copy} approach adds $k$ copies of the variables, one copy for each near-optimal solution desired, in an optimization model and adds constraints to enforce that the solutions differ \citep{Cameron.2021}.

\cite{Greistorfer.2008} compared the sequential and variable copy approaches for the problem of finding two diverse near-optimal solutions and found that the sequential approach usually required less computation time and yielded solutions that were nearly as good as the simultaneous approach. Further, population-based \textit{metaheuristics} are a natural approach for finding diverse sets of solutions because these algorithms operate on a set of solutions. For instance, \cite{glover2000scatter} used a scatter-search algorithm to find a diverse set of solutions to MIPs. While metaheuristics are often effective in practice, their main drawback is their inability to guarantee that the entire set of near-optimal solutions has been found.  Finally, \textit{two-phase} approaches use an oracle such as the {\sc OneTree} algorithm \citep{danna2007generating} to find a (not necessarily diverse) set of near-optimal solutions. Then the second phase inputs this set of near-optimal solutions and chooses a subset that maximizes diversity. Regarding the second phase, \cite{danna2009select} developed exact and heuristics algorithms to find a diverse subset. The work by \cite{Glover.1998} which proposed four different heuristics algorithms for generating the most diverse solutions from a larger set does not consider how close they are to the optimal objective value. \cite{kuo1993analyzing} proposed two different methods that use linear programming to select the most diverse solutions from a solution set. \cite{ schwind2020representative} developed methods for computing a small subset of solutions that represent the larger set using bi-objective optimization. In \cite{danna2007generating}, the authors found that the sequential and simultaneous methods do not scale as the number of solutions that are been generated increases. The approach developed in this study combines both phases of the two-phase approach, emphasizing diversity while searching for a near-optimal set.

\subsection{Contributions and Findings}
This study contributes to the existing literature on solution diversity in mixed integer programming by investigating how to obtain a diverse set of near-optimal solutions during the branch-and-bound search rather than during a post-processing step. We make the following contributions.\\

\noindent \textbf{Diversity of Solution Sets Computed by Existing Node Selection Rules.}  \ Existing branch-and-bound algorithms are not designed to find diverse sets of solutions, and it is unknown whether different algorithm configurations (e.g., node selection rules) influence the diversity of solution sets. To address this gap, we investigated the diversity of near-optimal sets generated using popular node selection rules within a branch-and-count algorithm \citep{Achterberg.2008}. We found that, while the effect of node selection rules on solution set diversity is problem specific, the \textit{best-first search} rule typically yields better diversity solutions overall in reasonable time.\\

\noindent\textbf{A New Node Selection Rule that Emphasizes Solution Set Diversity.} \ The current well-established methods for finding a diverse set of near-optimal solutions do so in a two-phase approach in which a set of near-optimal solutions is first computed and then a diverse subset is identified using a post-processing algorithm. In this study we describe a new approach that computes diverse sets of near-optimal solution sets within a single branch-and-bound search\footnote{to be clear, our method as well as existing methods also include a precomputation step in which the optimal objective value is computed. However, we do not count this step as a phase, when referring to the number of phases of our method or other methods.}. To accomplish this, we introduced and subsequently investigated a new parameterized node selection rule that selects new nodes in the tree based on a weighted average of the dual bound, partial solution diversity, and tree depth. We find that each of these three metrics are needed to find diverse solution sets.\\
    
\noindent\textbf{Tuned Parameters for Diversity-emphasizing Node Selection Rules.} \ Because the weighted-average metric used in our node selection rule requires parameters, we performed an experimental investigation to find the parameter combination that, on average, yields the highest solution diversity. To ensure that the identified parameter combination works well on new problems, we first tuned the parameters using a training set and then tested the resulting best parameter combinations on a test set. Results show that one particular parameter combination performs favorably well on both the training and tests sets over a variety of problem settings. Thus, our results indicate that our method generally performs well with a prescribed combination of parameters, alleviating the need for time-consuming parameter optimization and providing a set of default parameters that should work well on new problem not included in our test set.\\
    
\noindent\textbf{Benefits of Emphasizing Diversity in Node Selection.} \ Using 36 problem instances randomly selected from MIPLIB (2003, 2010 and 2017), we found that using our new diversity-emphasizing node selection rule results in solution sets that are up to 190\% more diverse than the {\sc OneTree} algorithm, a leading method for finding near-optimal sets currently implemented in CPLEX. Regarding runtime, we found that our method of emphasizing diversity within the node selection procedure runs within $\pm 15\%$ of the time used by regular node selection methods that do not emphasize diversity and results in a better tradeoff between runtime and solution diversity than competing approaches.
    

\subsection{Outline of the paper}
In the remainder of this article, we provide mathematical and algorithmic preliminaries in \S \ref{sec:2background}, followed by a description of the diversity-emphasizing node selection rules tested in this work in \S \ref{sec:3Diversity-EMphasizing}. Experimental results on the training and test set data is discussed in \S \ref{sec:4results}, while \S \ref{sec:example}
illustrates the use of {\sc DiversiTree} on real data. In \S \ref{sec:6Extensions}, we discuss extensions to the method outlined in this study, prior to concluding in \S \ref{sec:5conclusion}.

\section{Mathematical and Algorithmic Preliminaries}
\label{sec:2background}

We consider finding a diverse set of near-optimal solutions to the following problem:

\begin{equation}
\label{eqn:1MIP}
\begin{split}
     & z^* = \min_{x \in X}\, c^{T}x  \quad \text{ where} \\
    X &=\{x \in \mathbb{R}^{d} : Ax \geq b, x_{i} \in \mathbb{Z}, \forall i \in \mathbb{I} \subseteq \{1, \dots , d\}\}.
\end{split}
\end{equation}

\noindent Let $S_{q}=\{x\in X : c^T x \leq (1+q)z^*\}$ denote a set of $q$\%-optimal solutions to \eqref{eqn:1MIP}, $q \geq 0$. While mixed-integer programs are considered, diversity is only computed over binary integer variables. If $|S_{q}|$ is small, it may be sufficient to use a near-optimal solution generation algorithm such as {\sc OneTree} \citep{danna2007generating} to generate the entire set $S_{q}$ for presentation to a decision maker. In the case in which $|S_{q}|$ is large but not very large, not more than 1,000 elements, the following two-phase approach may be used. In the first phase, a \textit{solution generation} algorithm is used to obtain the complete set of near-optimal solutions $S_{q}$. Then in the second phase, use a \textit{diverse subset selection} algorithm~\cite[see, e.g.,][]{danna2009select} to find a small subset of $S_{q}$ of cardinality $p$, solving the following problem:

\begin{equation}
\label{eqn:maxdiversity}
\max_{S \subseteq S_{q},|S|=p} \mathcal{D}(S).
\end{equation}

\noindent where $\mathcal{D}(S)$ is a measure of the diversity of solution set $S$.

In many cases, however, the set $S_q$ can greatly vary in size, ranging from one or several elements, to perhaps 10,000 or more. In particular, for about half of the instances tested in \cite{danna2009select}, $|S_q| > 10,000$. For these instances, the authors limited the input to the second phase to a subset of $S_q$ consisting of the first 10,000 solutions obtained by the {\sc OneTree} algorithm \cite{danna2007generating}. That is, the first phase finds a subset $\bar{S}\subseteq S_q$ ($|\bar{S}| = 10,000$) without explicit consideration of diversity, while the second phase solves:

\begin{equation}
\label{eqn:maxdiversity}
\max_{S \subseteq \bar{S},|S|=p} \mathcal{D}(S).
\end{equation}

These problem instances with very large $S_q$ present two issues:

\begin{itemize}
    \item \textit{Computational}. These problems could not be solved to optimality in the second-phase of the two-phase approach due to memory limitations or exceeding a 10-day time limit \citep{danna2009select}. As a result, heuristics were employed. For smaller problems these heuristics appear to produce solutions that nearly maximize diversity. However, it is not known whether the heuristics produce good solutions to any of the larger problems.
    \item \textit{Solution quality}. Using the subset $\bar{S}$ as an input to the second phase rather than the complete set $S_q$ could result in a loss of diversity if there is insufficient diversity in the input subset to either of the heuristic or exact methods.
\end{itemize}

To address these issues, we seek to solve \eqref{eqn:maxdiversity} directly rather than via a two-phase approach. Specifically, we examine how to modify the exploration strategy of the solution generation phase to increase the diversity of the subset $\bar{S}$.

\subsection{Computing near-optimal sets}

Given a mixed integer programming problem, we assume the existence of an oracle capable of enumerating a set of all or a sufficiently high number of near-optimal solutions for the problem. We use the \textit{branch-and-count} method as the oracle due to its significant speed at enumerating near-optimal solutions. The branch-and-count method is implemented in SCIP \citep{GamrathEtal2020OO}, which provides access to, as well as modification privileges for, the branching tree and implementation methods of the oracle. The branch-and-count algorithm \cite{Achterberg.2008}, an extension of the branch-and-cut algorithm customized for detecting sets of near-optimal solutions, seeks to analyze all solution vectors contained in a subtree without complete enumeration. \textit{Infeasible subtrees}, subtrees in which all of the leaf nodes are infeasible, are straightforward to detect because the descendants of an infeasible node are also infeasible. Thus, branch-and-count focuses on detecting \textit{unrestricted subtrees} in which all of the leaf nodes are feasible. The authors show that a subtree rooted at a node is unrestricted if and only if all of the constraints at the node are locally redundant (i.e., satisfied by all possible variable assignments of values in the domain of that subtree).

In the context of finding sets of near-optimal solutions to an optimization problem, branch-and-count first requires knowledge of the optimal objective value $z^*$. Next the algorithm removes the objective value from the problem, adds the constraint $c^Tx\leq (1+q)z^*$, and solves the problem using an approach much like branch-and-cut but with unrestricted subtree detection. Thus, in this context an unrestricted subtree is one in which all of the leaf nodes in the subtree represent solutions in the near-optimal set. The pseudocode for branch-and-count is given in Algorithm \ref{alg:branchandCut}. 

To use this branch-and-count to generate the set $\bar{S}$, we first solve \eqref{eqn:1MIP} to find an optimal objective value $z^*$. Next, in line 1 of the algorithm, we pass the MIP problem and the number of near-optimal solutions requested ($p_1$) as input to the algorithm and add the constraint $c^{T}x \leq (1+q)z^{*} $ to the problem and solve it using the branch-and-count algorithm. We initialize a queue $\textit{Q}$ that holds nodes, each representing an MIP problem, that are available for exploration. We add the initial MIP to $\textit{Q}$ and also initialize set $\bar{S} = \emptyset$. While $\textit{Q}$ is not empty and the number of solutions in the set $\bar{S}$ is less than $p_1$, we select a node $i$ from $\textit{Q}$ using a node selection rule. If the problem associated with node $i$ is infeasible, we discard this node. If $i$ is determined to be unrestricted,  we collect solutions from $i$ and its subtree and add them to $\bar{S}$. Before discarding any node $i$, branching is performed, adding additional nodes to $\textit{Q}$ if the node is neither unrestricted nor infeasible. The entire process stops when either are no more nodes in $\textit{Q}$ to process or the cardinality of the set $\bar{S}$ equals $p_1$, at which point $\bar{S}$ is returned.

\begin{minipage}{.95\linewidth}
\begin{algorithm}[H]
\label{alg:example}
\caption{Branch and Count Pseudocode}
\label{alg:branchandCut}
\begin{algorithmic}[1]
\STATE \textbf{INPUT} $p_1$, \textit{MIP}
\STATE \textbf{ADD} constraint $c^{T}x \leq (1+q)z^{*} $ to \textit{MIP}
\STATE $\textit{Q} \gets \emptyset$
\STATE \textbf{ADD} \textit{MIP} to queue of active nodes \textit{Q}
\STATE $\bar{S} \gets \emptyset$
\WHILE{ \textit{Q} is not empty and $|\bar{S}| < p_1 $}
\STATE Use node selection rule to $dequeue$ node $i'$ from \textit{Q}
\IF{node $i'$ is \textit{unrestricted}}
\STATE Enumerate all child nodes of node $i$ and add their solutions to $\bar{S}$
\ELSIF{node $i'$ is \textit{infeasible}}
\STATE Discard node
\ELSE
\STATE Perform branching and add child nodes to queue \textit{Q}
\ENDIF
\ENDWHILE
\RETURN $\bar{S}$
\end{algorithmic}
\end{algorithm}
\end{minipage}

\vspace{1.5cm}

\subsection{Node Selection Rules.}
\label{sec:node-selection-rules}
During line 7 of the branch-and-count search outlined in Algorithm \ref{alg:branchandCut}, the node selection rule decides which node from the queue of active nodes is selected as the next node. Popular node selection rules include:
\begin{itemize}
    \item \textbf{Best-First Search (BestFS).} Selects the node with the best bound, that is, for minimization problems, select $i'$ as the next node according to
    
    \[i'\in \arg\min_{i\in Q} \left\{ LB_i \right\},\]
    
    where $LB_i$ is the lower bound for node $i$.
    
    \item \textbf{Depth First Search (DFS).} Nodes encountered as the branch and bound search tree is traversed are added to a queue and are selected in Last In First Out (LIFO) order.
    \item  \textbf{Breadth First Search (BrFS).} Nodes are added to a queue and processed using the First In First Out (FIFO) order.
    \item \textbf{Upper Confidence Bounds for Trees (UCT) \citep{GamrathEtal2020OO}.} Selects the next node $i'$ as a node with the best UCT\_score, that is:
    
     \[i'\in \arg\min_{i\in Q} \left\{ UCT\_Score_i \right\},\]
     
     $UCT\_score$ is calculated as:
    
    \[ UCT\_Score_i =  LB_i + \rho \frac{V_i}{v_i}  \]
    
    where $v_i$ and $V_i$ are the number of times the algorithm has visited node $i$ and its parent, respectively. $\rho$ is a weight parameter chosen by the user.
    
    \item \textbf{Hybrid Estimate (HE) \citep{GamrathEtal2020OO}.} Selects the next node $i'$ as a node having the best HE\_Score, i.e.:
    
    \[i'\in \arg\min_{i\in Q} \left\{ HE\_Score_i \right\},\]
    
    $HE\_Score$ is calculated as:
    
    \[ HE\_Score_i =  (1-\rho) LB_i + \rho \widehat{LB_i} \]
    
    $\widehat{LB_i}$ is the estimated value of the best feasible solution in subtree of node $i$ and $\rho$ is a user-defined weight parameter.

\end{itemize}

\subsection{Measuring the diversity of solutions}

\paragraph{} A number of metrics exist for measuring the diversity of a set of solutions. A good metric needs to be model agnostic and ideally scaled such that diversity scores given by the metric are easy to interpret. \cite{danna2009select} outlined three problem agnostic measures for the diversity of solutions: $DBin$, defined in more detail below and used in our tests, is scaled by the number of variables and solutions generated and considers just the binary variables; $DAll$ which considers all variable types; and $DCV$ which is the scaled version of $DAll$.\\

The $DBin$ metric (defined only on binary variables) is the average scaled Hamming distance between all pairs of solutions in a set $\solutionset$, that is,

\begin{equation}
\label{eqn:2Dbin}
\begin{split}
     DBin(\solutionset) & = \frac{2}{|\solutionset| (|\solutionset|-1)} \sum_{j=1}^{|\solutionset|} \sum_{k=j+1}^{|\solutionset|} ham(x^{(j)},x^{(k)}),  \\
\end{split}
\end{equation}\\

\noindent where $x^{(j)}$ is the $j^{th}$ solution generated by the oracle, and $ham()$ computes the Hamming distance between a pair of solutions, that is,

\begin{equation}
\label{eqn:3Dbin}
\begin{split}
     ham(x^{(j)},x^{(k)}) & = \frac{1}{|B|} \sum_{i \in B} | x_{i}^{(j)} - x_{i}^{(k)} |,  \\
\end{split}
\end{equation}\\

\noindent where $B$ is the set of binary variables and $|B|$ is the number of binary variables. An advantage of the $DBin$ metric is that it takes values between 0 and 1 inclusive and does not depend on the size of the solution set or the number of variables.

\section{Diversity-Emphasizing Node Selection rules}
\label{sec:3Diversity-EMphasizing}
Within the branch-and-count algorithm \citep{Achterberg.2008}, we investigated several variants of the best-first search (BestFS) node selection rule that consider solution set diversity when selecting the next node to evaluate. We focused on BestFS because, as our results in \S \ref{sec:4results} show, it generated the most diverse solution sets in reasonable time when compared with other well-known node selection rules (such as DFS) when diversity was not considered in the node selection task. What follows is a description of each of the custom node selection rules tested in this work.

\subsection{Diverse-BFS (D-BFS($\alpha$))}
\label{sec:3.1 AlphaOnly}
The Diverse-BFS (D-BFS $\alpha$) node selection rule considers both the lower bound of a node as well as the diversity of the node with respect to other solutions already in the near-optimal set. That is, this rule inputs a set of open nodes $O$ and selects the next node $i'$ according to:

\[ i'\in \arg\min_{i\in O} \{  (1-\alpha) \mathcal{L}_i  +  \alpha D_i  \}, \]

\noindent where $\alpha\in [0,1]$ is a parameter that trades off the bound of node $i$ against its diversity score $D_i$ and $\mathcal{L}_i$ is a scaled lower bound of node $i$, that is,

\[ \mathcal{L}_i = \frac{ LB_i - \min_{j\in O} LB_j }{\max_{j\in O} LB_j - \min_{j\in O} LB_j }. \]

The lower bound is scaled to $[0,1]$, commensurate with the diversity score $D_i \in [0,1]$. The value of $D_i$ represents the \textit{partial diversity} of node $i$ with respect to the current solution set $\bar{\solutionset}$. We use the term \textit{partial} because it is computed based only on the binary variables that have been fixed at a particular node in the branching tree.

\subsection{Diverse-BFS with tree depth (D-BFS($\alpha,\beta$))}
\label{sec:3.1.2 AlphaBetaOnly}

The D-BFS($\alpha,\beta$) node selection rule considers the lower bound, the diversity and the depth of a node with respect to other solutions in the near-optimal set. This rule inputs a set of open nodes $O$  and selects the next node $i'$ according to:

\[ i'\in \arg\min_{i\in O} \left\{ (1-\alpha-\beta) \mathcal{L}_i + \alpha D_i + \beta H_i \right\}, \]

\noindent where $(1-\alpha-\beta),\alpha, \beta \in [0,1]$ form a convex combination and are the parameters that control the weight of the scaled lower bound of node $i,  \mathcal{L}_i  $, the diversity score $D_i$, and the scaled depth of node $i, H_i $. Like $\mathcal{L}_i$, $H_i$ is scaled as:

\[ H_i = \frac{ Depth_i - MinPlungeDepth }{MaxPlungeDepth - MinPlungeDepth}, \]

\noindent where \textit{MinPlungeDepth} and \textit{MaxPlungeDepth} are set at the beginning of the computation and $Depth_i$ is the depth of node $i$ in the tree.

\subsection{Diverse-BFS with solution cutoff (D-BFS($\alpha, s$))}
\label{sec:3.1.3 AlphaSolCutOff}

The D-BFS($\alpha,s$) rule considers only the lower bound of a node until it has generated a small set of up to $s$ solutions prior to incorporating  the diversity of a node with respect to the solutions already in the near-optimal set.  This rule selects the next node $i'$ according to: 

\[ i' \in \begin{cases}
 \arg\min_{i\in O} \left\{ \mathcal{L}_i \right\} & \text{number of solutions found so far} $\textless s$, \\
 \arg\min_{i\in O} \left\{ ( 1 - \alpha) \mathcal{L}_i + \alpha D_i \right\} & \text{otherwise,}
\end{cases} 
\]

\noindent where $s$ is the solution cutoff parameter, that is, the number of solutions that must be accumulated before diversity is considered in node selection and $\alpha\in [0,1]$ is a parameter that trades off the bound of node $i$ against its diversity score $D_i$.

\subsection{Diverse-BFS with depth cutoff (D-BFS($\alpha, d$))}
\label{sec:3.1.4 AlphaBetaDepthCutOff}

The D-BFS($\alpha, d$) rule considers only the lower bound of a node until the depth of the active node reaches a depth $d$ prior to also considering the diversity of the node with respect to the solutions already in the near-optimal set. This rule selects the next node $i'$ according to:

\[ i' \in  \begin{cases}
 \arg\min_{i\in O} \left\{ \mathcal{L}_i \right\} & \text{ depth of nodes in current iteration $< d$,} \\
 \arg\min_{i\in O} \left\{ (1-\alpha) \mathcal{L}_i +  \alpha D_i \right\} & \text{ otherwise, }
\end{cases} \]

\noindent where $d$ is the depth cutoff parameter, that is, the depth that must be reached before diversity is considered in the node selection and $\alpha\in [0,1]$ is a parameter that trades off the bound of node $i$ against its diversity score $D_i$. Diversity is only triggered upon reaching a depth of $d$ or greater.

\subsection{{\sc DiversiTree} - Diverse-BFS with solution cutoff and tree depth (D-BFS($ \alpha, \beta, s$))}
\label{sec:3.5 AlphaBetaDepthSolCutOff}

{\sc DiversiTree} selects the next node $i'$ according to:

\[ i\in \begin{cases}
 \arg\min_{i\in O} \left\{ \mathcal{L}_i \right\} &  \text{ number of solutions found so far $< s$,} \\
 \arg\min_{i\in O} \left\{ ( 1 - \alpha - \beta) \mathcal{L}_i + \alpha D_i +  \beta H_i \right\} & \text{ otherwise, }
\end{cases} \]

\noindent where $\alpha, \beta, s$ are parameters as defined in previous sections.

\subsection{Other Diverse-BFS methods tested}
\label{sec:3.1.5 SingleDiversityDepthSolCutOff}
We also tested several other methods for emphasizing solution set diversity. However, these additional methods were not as effective as the methods described in \S\ref{sec:3.1 AlphaOnly} - \S\ref{sec:3.5 AlphaBetaDepthSolCutOff} above. They are:

\begin{enumerate}
    
    \item Using the minimum of diversity and depth, that is, select next node $i'$ according to: 
    
    \[ i'\in \arg\min_{i\in O} \left\{ (1-\alpha) \mathcal{L}_i + \alpha (min(D_i,H_i)) \right\}. \] 
    
    \item Using the maximum of diversity and depth, that is, select next node $i'$ according to: 
    
    \[ i'\in \arg\min_{i\in O} \left\{ (1-\alpha) \mathcal{L}_i + \alpha( max(D_i,H_i) \right\}. \] 
    
    \item Using the product of diversity and depth, that is, select next node $i'$ according to: 
    
    \[ i'\in \arg\min_{i\in O} \left\{ ( 1-\alpha) \mathcal{L}_i + \alpha D_i H_i \right\}. \] 
    
\end{enumerate}

These node selection rules were also tested with a nonzero solution cutoff parameter $s$, but were still not effective.

\section{Computational Experiments.}
\label{sec:4results}
To measure the effect of using diversity-emphasizing node selection rules on solution set diversity, we ran several sets of experiments on selected problems from MIPLIB \citep{KochEtAl2011,BixbyCeriaMcZealSavelsbergh1998,MIPLIB2017} using several different node selection rules, including the customized ones described in \S \ref{sec:3Diversity-EMphasizing}. We compared our approach with two state-the-art methods: {\sc OneTree} \citep{danna2007generating} and {\sc Branch-And-Count} \citep{Achterberg.2008}.

Experiments were run to answer the following questions:
\begin{enumerate}
    \item[\textit{Research Question 1}:] Among common node selection rules (such as BestFS and DFS), do some produce a more diverse set of near-optimal solutions than others? \textbf{\S \ref{sec:diversity-of-common-node-selection}}
    \item[\textit{Research Question 2}:] What are the best parameters to use for the parameterized diversity-emphasizing node selection rules presented in \S \ref{sec:3Diversity-EMphasizing}? When using the best parameters, do the diversity-emphasizing node selection rules compute solution sets that are more diverse than those computed by competing approaches? \textbf{\S\ref{sec:parameter-optimization}}
    \item[\textit{Research Question 3}:] What parameters should be used when using diversity-emphasizing node selection rules on a new problem? \textbf{\S\ref{sec:TestSet}}
\end{enumerate}

\subsection{Experimental Setup}
All code was implemented in C++ using SCIP Optimization suite 7.0 \citep{GamrathEtal2020OO} and run on a server running Intel Xeon processors with sixteen cores and thirty two GB of memory. Apart from SCIP currently being one of the fastest non-commercial solvers for mixed integer programming (MIP) and mixed integer nonlinear programming, it provides a convenient way to use custom node selection rules. We evaluated our methods against the state-of-the-art {\sc OneTree} method, implemented in GAMS-CPLEX.

\subsection{Diversity of common node-selection rules}
\label{sec:diversity-of-common-node-selection}
We first examined the diversity produced by common node selection rules. For this initial set of experiments, we selected seven problems from MIPLIB that had greater than 10,000 solutions within 1\% of optimality (see Table \ref{fig:problemsSolved2} in Appendix \ref{Sec:trainingSet}). 

We used similar settings on the common node selection rules listed in \S \ref{sec:node-selection-rules} to request the generation of sets with sizes ranging from 50 to 2,000 near-optimal solutions, and computed the diversity. Figure \ref{fig:mainScipDiversity} below shows the $DBin$ diversity scores achieved by these common node selection rules for different solution set sizes ($p_1$). We labeled the different node selection rules with a ``BC'' prefix to indicate the use of branch-and-count to generate the solution sets. As shown in Figure \ref{fig:mainScipDiversity}, the best-first search (BCBFS) method finds sets with the greatest diversity in 5 out of 7 test problems. The upper confidence bounds for trees (BCUCT) node selection rule had the best performance in 2 out of the 7 test problems and achieved a similar diversity value to hybrid estimate (BCHE) in all other cases. The depth first search (BCDFS) was outperformed in terms of diversity in all of the instances.

The plots also indicate that diversity (averaged for each problem instance) tends to start low and become incrementally higher as the number of requested solutions increases. We hypothesize that this may be due to a situation where as more solutions are generated, more variables are fixed or modified; increasing the Hamming distance from the first solution found. However, this trend did not hold for all of the individual problem instances.

\begin{figure}[H]
    \centering
    \includegraphics[width=\linewidth]{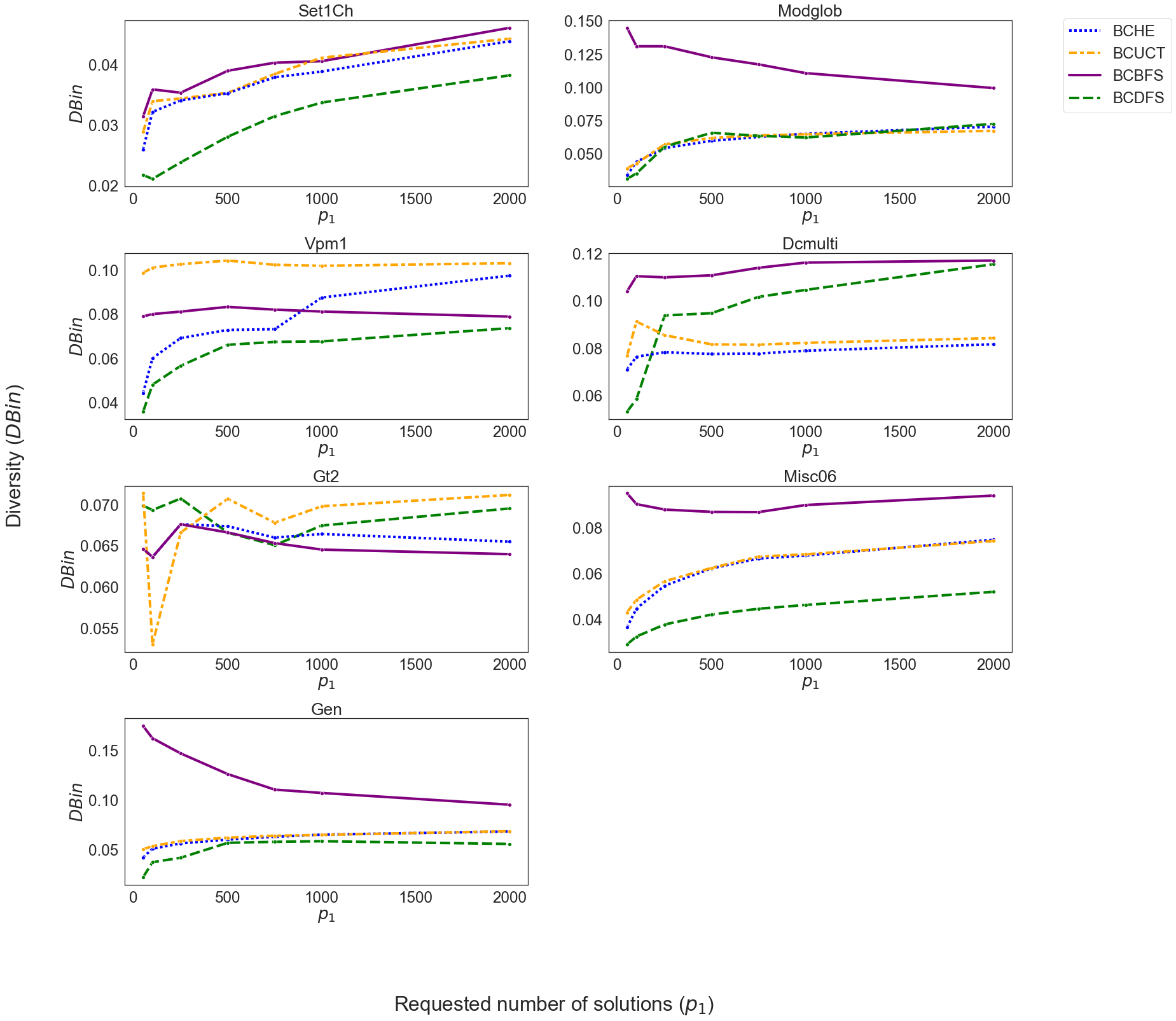}
    \captionsetup{margin=2cm,format=plain, font=small, singlelinecheck=false}
    \caption{The final $DBin$ diversity score achieved by common node selection rules (BestFS, DFS, UCT and HE) available on SCIP. We prefix each node selection rule name with "BC" to indicate the use of the "branch-and-count" method. The BestFS node selection rule resulted in greater diversity for a majority of problems.}
    \label{fig:mainScipDiversity}
\end{figure}

The superior performance of BestFS over other common rules suggested its continued use for comparison in the remainder of our experiments.

\subsection{Parameter optimization for diversity-emphasizing rules: training set}
\label{sec:parameter-optimization}
The most general diversity-emphasizing node selection rules have up to three parameters to be tuned: $\alpha$ controls the emphasis on diversity for the selected node, $\beta$ controls the emphasis on depth within the search tree for the selected node, and $S$ controls the emphasis on the number of solutions generated before employing the $\alpha$ and $\beta$ values in the node selection process.
We used a grid search to tune the parameters and find the best-performing values. Toward this end, we used several problem instances from MIPLIB \citep{MIPLIB2017, KochEtAl2011, BixbyCeriaMcZealSavelsbergh1998} and randomly divided the set of problems into a training and testing set using a 75:25 split. We used the training set problems to find the best parameters and then tested the performance of these parameters on the testing set problems. Using a grid-search on a 20$\times$20$\times$20 grid over $\alpha, \beta$, and $S$ on $[0,1]$ with increment 0.01 on $[0,0.09]$ and 0.1 on $[0.1,1]$, we tested different parameters values across all training set problems shown in Table \ref{fig:problemsSolved2} in Appendix \ref{Sec:trainingSet} for different numbers (10, 50, 100, 200 and 1000) of requested near optimal solutions $p_1$ and values of $q$ (\% near optimal) in the range $[0.01,0.1]$ with increment 0.01. As a result, we obtained for each instance the best performing values of $\alpha$, $\beta$, and $S$ over every $p_1$ and $q$. 

It is computationally impractical to run a grid search for every new problem instance prior to applying our node selection rule. We therefore attempted to identify patterns in the mapping between problems and optimal parameter values, so as to allow the identification of a smaller number of parameter settings with good performance. Thus, our next step was to determine whether we could group problems together based on their optimal parameter settings. Toward this end, we ran a standard hierarchical clustering algorithm available in Python's Scikit-Learn package \citep{scikit-learn} to cluster the problems into groups based on their best parameter settings, that is, the settings that yielded the highest $DBin$ scores for that problem. When the number of requested solutions is small (that is, 10), the clustering algorithm found the following four general groups:

\begin{enumerate}
    \item High $\alpha$, High $S$, Low $\beta$ (\textbf{HHL}): $\alpha \geq0.9$, $S\geq0.7$, $\beta\leq0.2$.
    
    \item High $\alpha$, Low $S$, Low $\beta$ (\textbf{HLL}): $\alpha \geq0.9$, $S \leq0.2$, $\beta \leq0.2$.
    
    \item Low $\alpha$, Low $S$, High $\beta$ (\textbf{LLH}): $\alpha \leq 0.2$, $S \leq 0.2$, $\beta \geq 0.8$.
    
    \item Low $\alpha$, High $S$, High $\beta$ (\textbf{LHH}): $\alpha \leq 0.2$, $S \geq 0.7$, $\beta \geq 0.8$.
\end{enumerate}

In the HHL group, diversity is emphasized heavily ($\alpha \geq 0.90$), but only after a large number of solutions have been accumulated ($S\geq 0.70$). In the HLL group, diversity is also heavily emphasized, but starting after a moderate number of solutions have been found ($S \leq 0.20$). The LLH group tends to select solutions at greater depths in the tree ($\beta\geq0.80$) after a small number of solutions have been accumulated ($S\leq0.20$). Finally, the LHH group emphasizes diversity lightly ($\alpha \leq 0.20$) and mostly selects solutions at greater depths in the tree ($\beta\geq0.80$) after a large number of solutions have been accumulated ($S\geq0.70$). The structure of these four groups indicates that our method is most effective when we generate a small number of seed solutions and then emphasize diversity at greater depths in the tree.

To assess the efficacy of {\sc DiversiTree}, we then selected a single parameter setting for each of the four groups, as shown in Table \ref{fig:parameterValues}. Using the data from the grid search, we took the weighted average of the settings in a group as $G_1$ and the settings that occurred the most frequently in that group as $G_2$. We computed the diversity of the training set problems using $G_1$ and $G_2$ separately and selected either $G_1$ or $G_2$ as the best setting for the group, depending on which of the two produced the best diversity. We found that these parameter settings worked well for different values of $p_1$ and $q$.

{\small
\centering

\begin{xltabular}{\linewidth}{@{  }L @{  }L @{  }L @{  }L @{  }L}
\caption{The $\alpha,\beta$ and $S$ parameter settings used in both training and testing.}
\label{fig:parameterValues}\\
   \hline
  Number of \newline solutions \newline requested  & HHL & HLL &  LLH  & LHH  \\
  \hline
   10, 50, 100, 200, 1000  & $\alpha$ :0.94, \newline $\beta$ :0.06, \newline $s$ :0.80 & $\alpha$ :0.95, \newline $\beta$ :0.06, \newline $s$ :0.20 & $\alpha$ :0.01, \newline $\beta$ :0.99, \newline $s$ :0.05 & $\alpha$ :0.18, \newline $\beta$ :0.8, \newline $s$ :0.70 \\
   \hline
\end{xltabular}
}

For each problem instance, we then ran {\sc DiversiTree} using the parameters from its assigned group from the clustering algorithm (see Figure \ref{fig:parameterGroups} in the Appendix). The parameter settings in Table \ref{fig:parameterValues} were used for all values of $p_1$ and $q$. Specifically, we used the following two-phase approach. In phase one, we generate a larger solution set $p_1$ with 10, 50, 100, 200, and 1000 solutions. In phase two, we use a subset selection method (similar to local search algorithm in \cite{danna2009select}) to select $p=10$ diverse solutions---a number that seems reasonable to present to a decision maker. We ran this process for values of $q$ from 1\% to 10\% across all the problems in the training set. We compared the performance of our method with two state-of-the-art methods: with the BCBFS rule, and with the {\sc OneTree} method results reported in \cite{danna2009select}. These competing methods were used in phase one of the two-phase approach, with the subset selection method being employed in phase two. 

\begin{figure}[H]
     \centering
     \begin{subfigure}[b]{\textwidth}
         \centering
         \includegraphics[width=\textwidth]{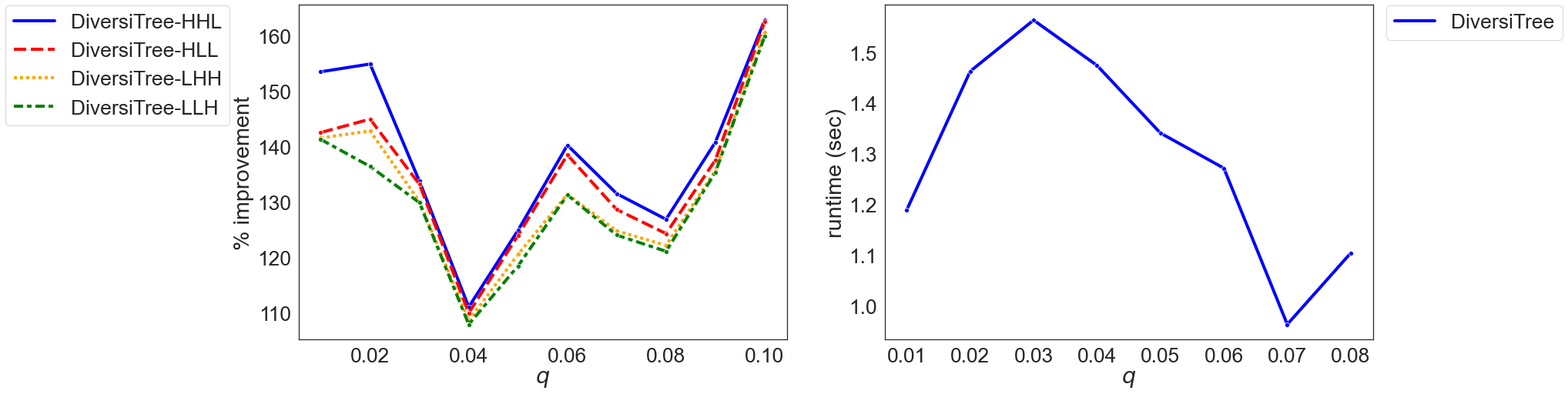}
         \caption{$p_1 = 10$, $p = 10$}
         \label{fig:p110}
     \end{subfigure}\\
      \begin{subfigure}[b]{\textwidth}
         \centering
         \includegraphics[width=\textwidth]{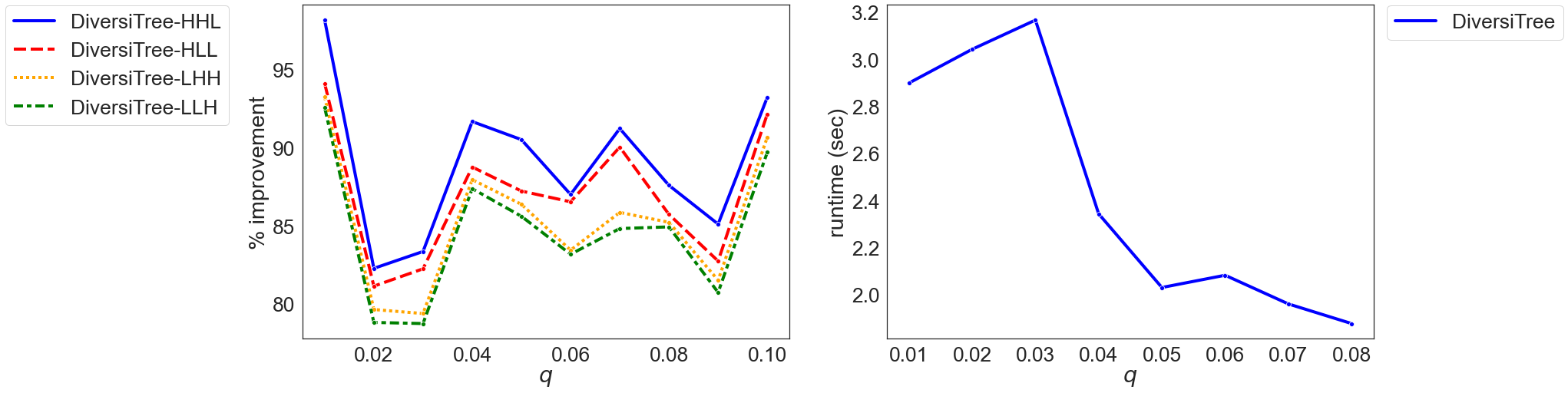}
         \caption{$p_1 = 50$, $p=10$}
         \label{fig:p150}
     \end{subfigure}\\
    \begin{subfigure}[b]{\textwidth}
         \centering
         \includegraphics[width=\textwidth]{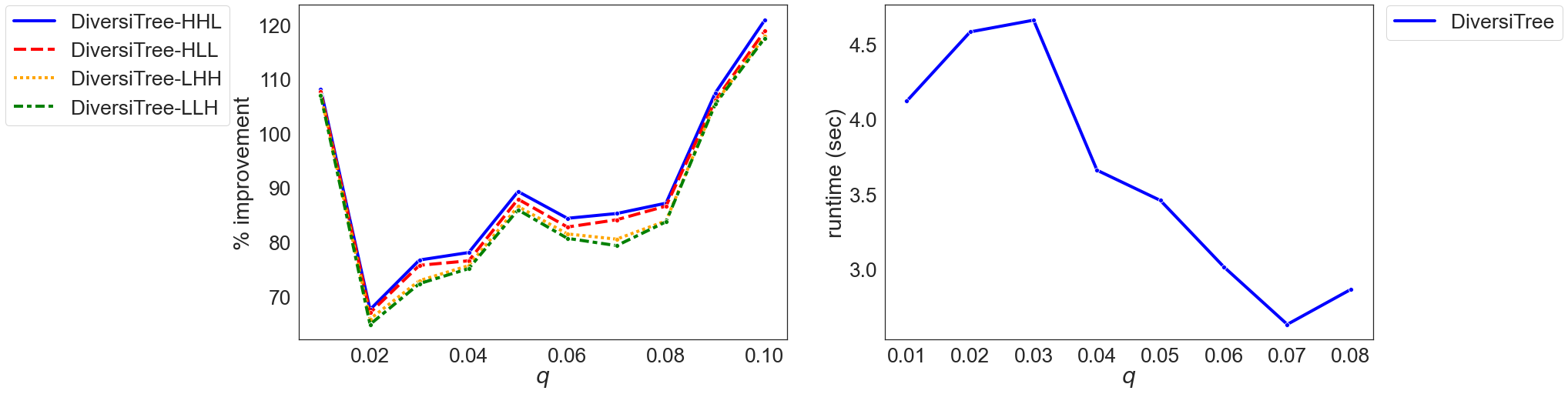}
         \caption{$p_1=100$, $p=10$}
         \label{fig:p1100}
     \end{subfigure}\\

    \captionsetup{format=plain, font=small, labelfont=bf,justification=raggedright,singlelinecheck=false,font=small}
    
    \caption{This figure shows plots for $p_1 \leq 100$. The plots on the left shows the average percent improvement on diversity ($DBin$) achieved by different parameter group settings of {\sc DiversiTree} over {\sc OneTree} aggregated across all problems in the training set.  The run times are shown in the plots on the right. The runtime for {\sc DiversiTree} were similar for the four different parameter groups, hence this figure shows the average runtime over the four groups.}
    
    \label{fig:trainingSetPlots1}
\end{figure}

\begin{figure}[H]
     \centering
     
          \begin{subfigure}[b]{\textwidth}
         \centering
         \includegraphics[width=\textwidth]{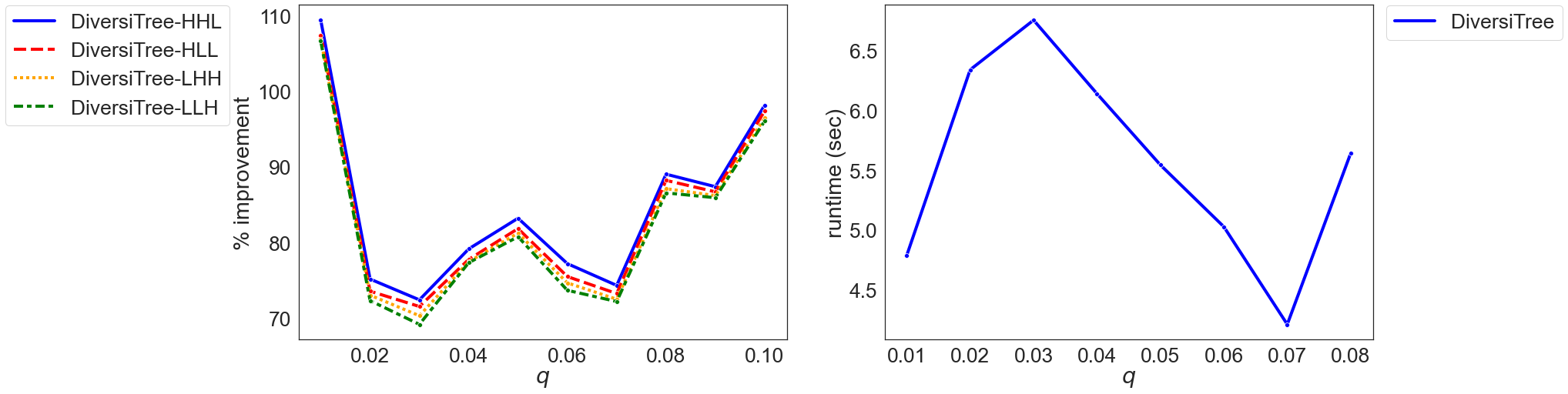}
         \caption{$p_1=200, p=10$}
         \label{fig:p1200}
     \end{subfigure}\\
        
      \begin{subfigure}[b]{\textwidth}
         \centering
         \includegraphics[width=\textwidth]{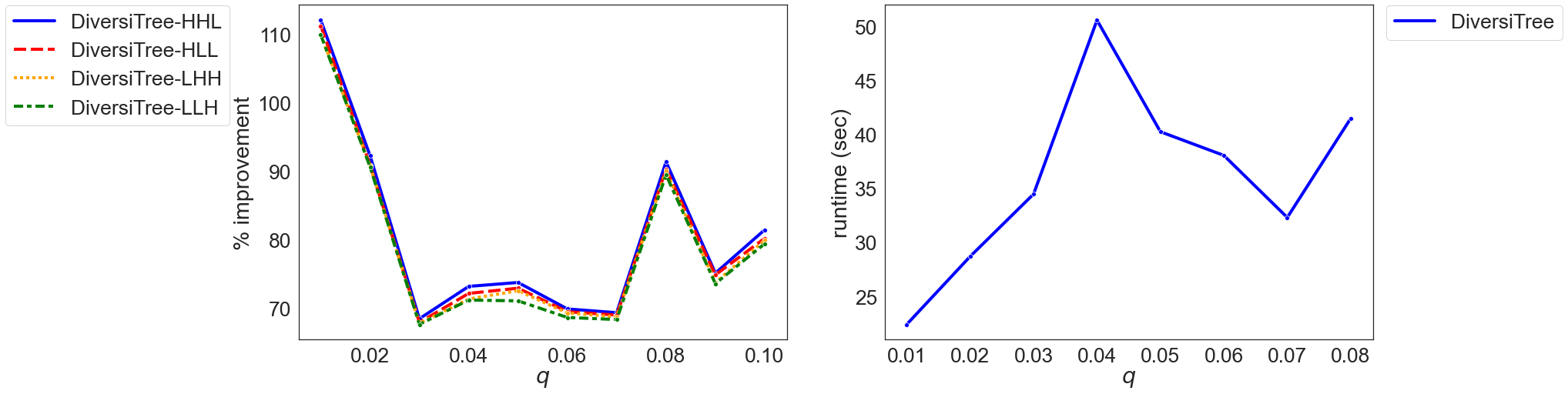}
         \caption{$p_1=1000, p=10$}
         \label{fig:p11000}
     \end{subfigure}\\
         
    \captionsetup{format=plain, font=small, labelfont=bf,justification=raggedright,singlelinecheck=false,font=small}
        
    \caption{This figure shows plots for $p_1 \geq 200$. The plots on the left shows the average percent improvement on diversity ($DBin$) achieved by different parameter group settings of {\sc DiversiTree} over {\sc OneTree} aggregated across all problems in the training set.  The run times are shown in the plots on the right. The runtime for {\sc DiversiTree} were similar for the four different parameter groups, hence this figure shows the average runtime over the four groups.}
        
    \label{fig:trainingSetPlots2}
\end{figure}

In Figure \ref{fig:trainingSetPlots1} and \ref{fig:trainingSetPlots2} we plot the percent improvement of the different groups of {\sc DiversiTree} over {\sc OneTree} for values of $q$ from $1\%$ to $10\%$, and $p=10$ for different $p_1$ values. In Figure \ref{fig:trainingSetPlots1}, we show plots for $p_1 \leq 100$ and in Figure \ref{fig:trainingSetPlots2}, we show plots for $p_1 \geq 200$. As the figure shows, {\sc DiversiTree} achieves an improvement in diversity over {\sc OneTree} of up to 160\% and at least 60\%, irrespective of the four parameter settings used. Beyond outperforming {\sc OneTree}, all the parameter groups also give similar performance on average with no more than $15\%$ gap between any group. Among the {\sc DiversiTree} results, parameter groups HHL and HLL appear to perform best. Thus, we recommend that when using our method on a new problem either the HHL ($\alpha=0.94, \beta=0.06,s=0.80$) of HLL ($\alpha=0.95, \beta=0.06,s=0.20$) parameter settings should be used. In the next section we test the HHL parameter setting on a set of new problems that were not used for parameter tuning.

\vspace{0.5cm}
\setlength{\tabcolsep}{18pt}

\subsection{Performance of tuned parameter groups on test set of problems}
\label{sec:TestSet}
The results of the previous section raise the question of how parameters should be tuned on a new problem. To answer this question, we tested our method using the HHL parameter setting, which was recommended in the previous section. We tested our method with the HHL setting on a test set of problems not used in parameter tuning (see Table \ref{fig:problemsSolved3} in Appendix \ref{Sec:testingSet}). We ran the same \textit{two-phase} approach for diverse, near-optimal solution generation as we did in the training phase. We ran this process for $p_1 \in \{ 10, 50, 100, 200, 1000 \}$ and $q$ from 1\% to 10\%. We kept the size of the final solution set to be $p=10$. Although our method performed better than {\sc BCBFS}\footnote{In using BCBFS, we found that the default parameters on SCIP (i.e., $ MINPLUNGEDEPTH=-1 $, $ MAXPLUNGEDEPTH=-1 $, and $ MAXPLUNGEQUOT=0.25 $) performed poorly. Thus, we manually tuned these settings to improve the diversity obtained by BCBFS. These tuned parameter settings were used in the results shown in the remainder of the paper.} and {\sc OneTree} for all four of the parameter settings, we only present results for the recommend HHL setting. 

\begin{figure}[H]
     \centering
     
     \begin{subfigure}[b]{\textwidth}
         \centering
         \includegraphics[scale=0.22]{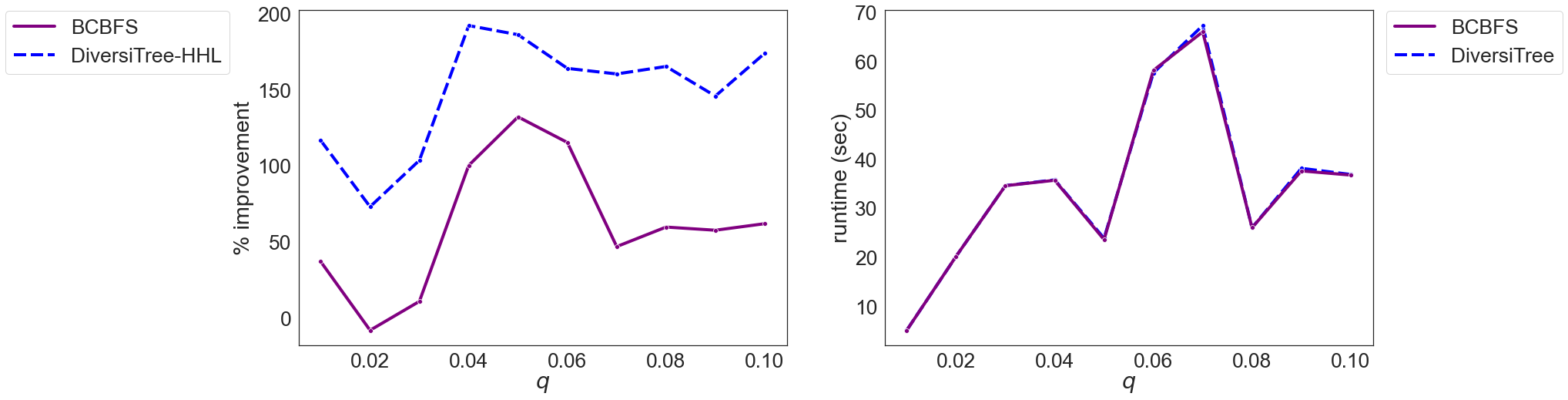}
         \caption{$p_1 = 10, p = 10$}
         \label{fig:p110}
     \end{subfigure}\\
         \begin{subfigure}[b]{\textwidth}
         \centering
         \includegraphics[scale=0.22]{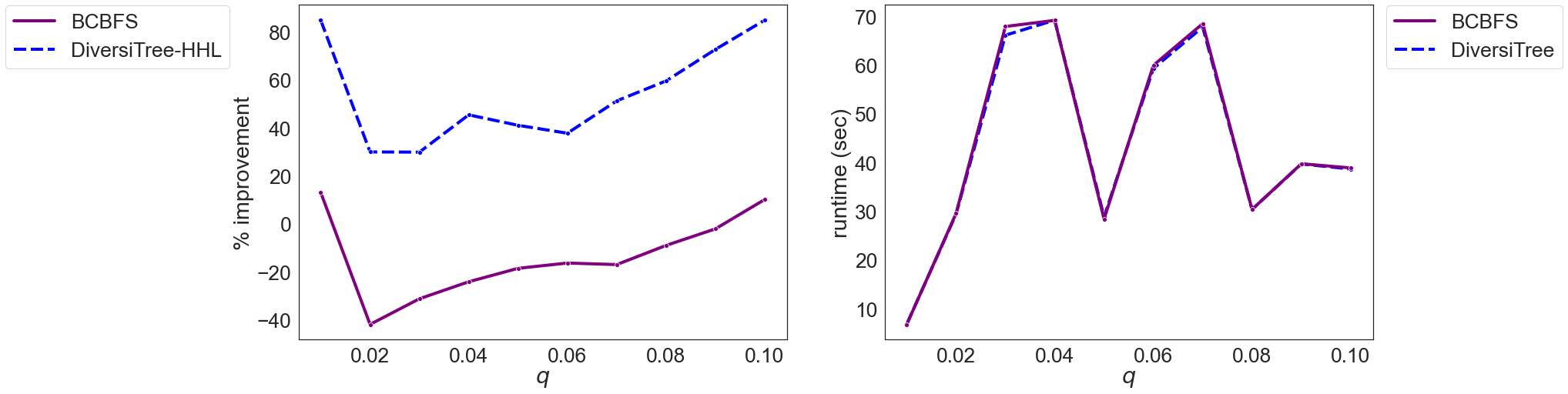}
         \caption{$p_1 = 50, p=10$}
         \label{fig:p150}
     \end{subfigure}\\
     \begin{subfigure}[b]{\textwidth}
         \centering
         \includegraphics[scale=0.22]{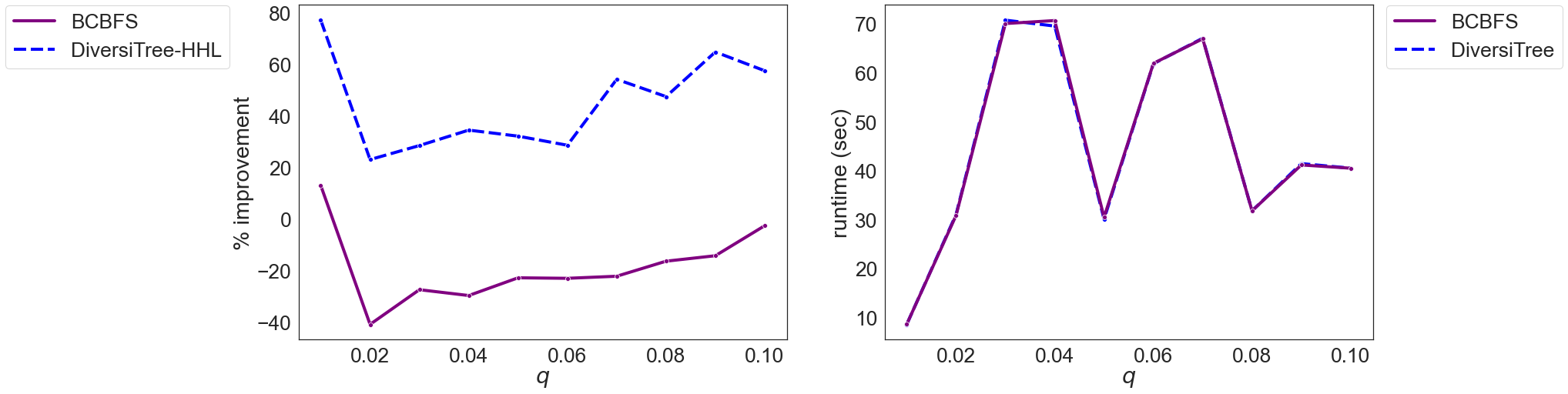}
         \caption{$p_1=100, p=10$}
         \label{fig:p1100}
     \end{subfigure}\\
    
    \caption{This figure shows plots for $p_1 \leq 100$. Plots on the left show the percent improvement on diversity ($DBin$) achieved by {\sc DiversiTree} and {\sc BCBFS} on the problems in the test set using the parameter setting from group HHL shown in Table \ref{fig:parameterValues}. The plots on the right show the runtime. The runtime for {\sc DiversiTree} were similar for the four different parameter groups, hence this figure shows the average runtime over the four groups.}
    
    \label{fig:testingSetPlots1}
\end{figure}

\begin{figure}[H]
     \centering
    
    \begin{subfigure}[b]{\textwidth}
         \centering
         \includegraphics[scale=0.22]{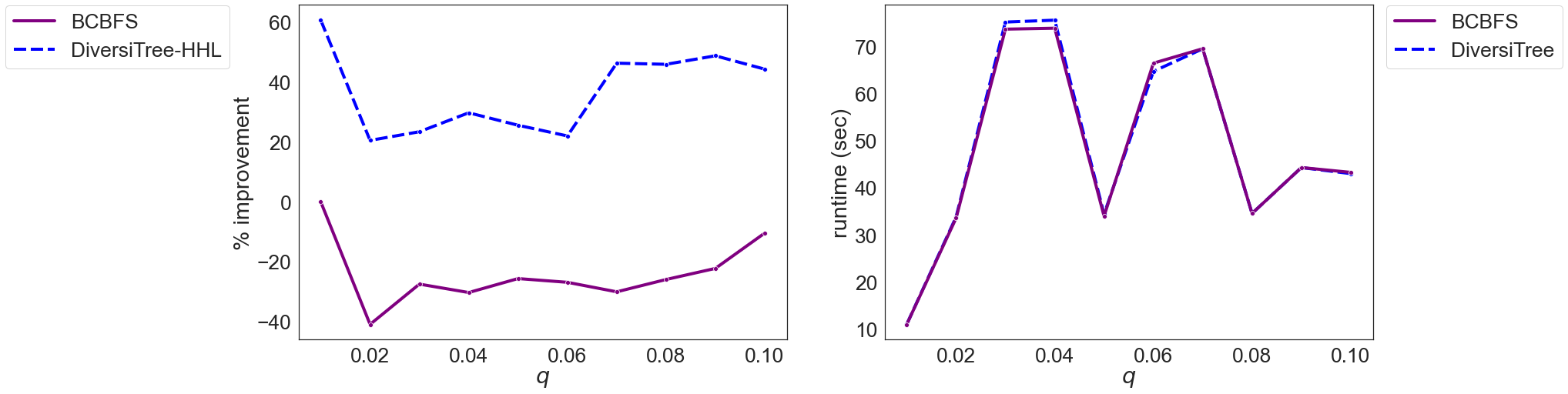}
         \caption{$p_1=200, p=10$}
         \label{fig:p1200}
     \end{subfigure}\\
     
         \begin{subfigure}[b]{\textwidth}
         \centering
         \includegraphics[scale=0.22]{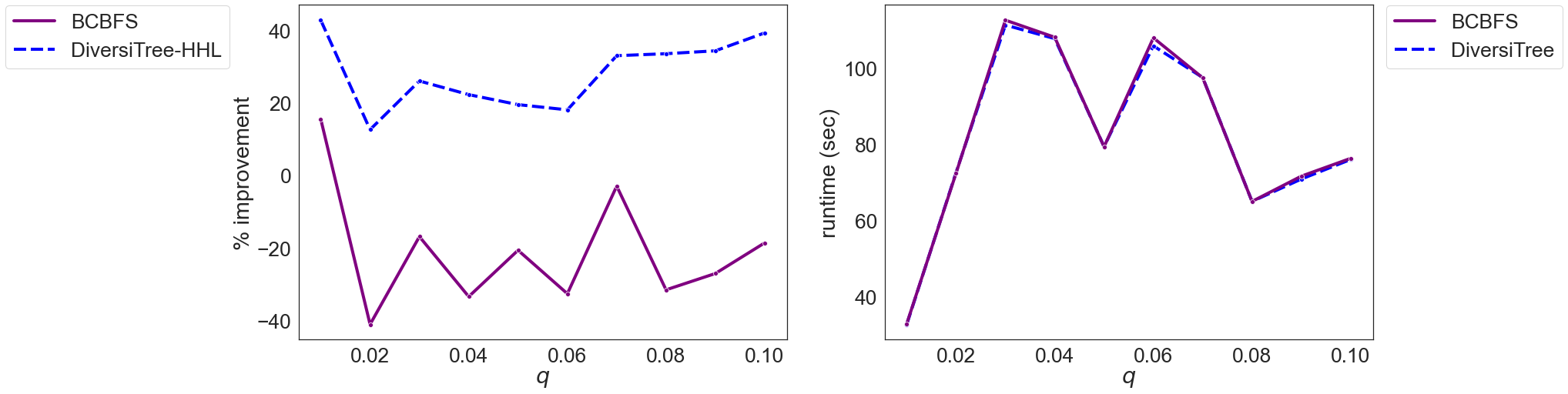}
         \caption{$p_1=1000, p=10$}
         \label{fig:p11000}
     \end{subfigure}\\
    
    \caption{This figure shows plots for $p_1 \geq 200$. Plots on the left show the percent improvement on diversity ($DBin$) achieved by {\sc DiversiTree} and {\sc BCBFS} on the problems in the test set using the parameter setting from group HHL shown in Table \ref{fig:parameterValues}. The plots on the right show the runtime. The runtime for {\sc DiversiTree} were similar for the four different parameter groups, hence this figure shows the average runtime over the four groups.}
    
    \label{fig:testingSetPlots2}
\end{figure}

We used the HHL parameter setting on the test data and the results achieved are shown in Figure \ref{fig:testingSetPlots1} and \ref{fig:testingSetPlots2}, which displays results similar to those in Figures \ref{fig:trainingSetPlots1} and \ref{fig:trainingSetPlots2} for the training set. {\sc DiversiTree} generated solution sets with improvement over {\sc OneTree} between 12\% and 190\%. {\sc DiversiTree} also significantly outperforms BCBFS in terms of diversity for all values of $p_1$, as BCBFS  generated solution sets with improvement of up to 130\% and in some instances performed worse than {\sc OneTree} by more than 35\%. In terms of runtime, {\sc DiversiTree} achieves similar runtime as BCBFS, although the {\sc OneTree} method runs significantly faster than both methods, which is unsurprising given that while {\sc DiversiTree} explicitly searches for solution, {\sc OneTree} only focuses on near-optimality. Unlike the training set data where parameter groups HHL and HLL dominated the {\sc DiversiTree} results, the parameter groups LHH and LLH dominated the diversity results generated in the test set.

In practice, a decision-maker seeking a diverse set of near-optimal solutions would need to define $p$ (the number of near-optimal solutions), $q\%$ (a bound on how far these solutions may be from optimality), and select a setting in any of the groups we specified in {\S \ref{sec:parameter-optimization}} (without need for parameter tuning); $p$ near-optimal solutions can then be directly generated within $q\%$ of the optimal, or a \textit{two-phase} approach can be used to generate the solution set. The result of our tests suggest that the solution set generated would be no worse than BCBFS and would be significantly better than the state-of-the-art {\sc OneTree} method.

\vspace{1cm}

\section{Example: Railway Timetabling Problem}
\label{sec:example}

To provide a concrete example, we tested {\sc DiversiTree} on a public transport scheduling problem within the MIPLIB2017 set \citep{MIPLIB2017}. Details about the problem formulation and variables is available in \cite{timtab1}. \cite{liebchen2002cyclic} discuss the general problem formulation. The problem models the cyclic railway timetabling problem where we have information about a railway network in a graph representing its infrastructure and traffic line. Each traffic line is operated every $T$ time units and the goal of the problem is to determine periodic departure times within the interval $[0,T)$ at every stop of every line. The problem has 397 variables; 77 of them are binary variables, 94 integer variables and 226 continous variables. The final objective is to obtain an arrival/departure timetable with minimal passenger and vehicle waiting times. The problem is represented as a graph $G=(N,A,l,u)$ with a set of nodes $N$ representing a set of events, and arcs $A$ representing set of trains. The weights on the arcs represent the time when event $v_i \in N$ occurs and $[l,u]$ the allowable time interval for this event. The events are represented as triplets (train,node,arrival) or (train,node,departure) and the binary variable $b_{ij}$ is 1 if arc $(i,j)$ is selected. The model is mainly based on \textit{periodic constraints}, which relate the arrival and departure time variables through the time window $[l,u]$ within which the event $i$ must occur.

For problems such as these, the solutions provided by the MIP model will be reviewed by a decision maker prior to implementation. Thus, in this context it can be useful to provide the decision maker with a set of near-optimal solutions from which to choose. This will allow the decision maker to consider not only the cost of the solution but also other factors such as variability in estimated demands, crew shortages, behavior of new equipment, or propagated delays, without having to re-solve the problem. For $q=1\%$, the problem had 181 near-optimal solutions, and when $q$ increased to 3\%, the number of near-optimal solutions increased to 6,264.

We ran the {\sc DiversiTree} and {\sc OneTree} algorithms on this problem for $p_1=10, p=10$ and $q=3\%$. {\sc DiversiTree} computed a solution set with a $DBin$ value of 0.299, while {\sc OneTree}'s set has a $DBin$ value of 0.044. A further review of the generated solutions shows that of the 77 binary variables, {\sc OneTree} gave a solution set in which 68 of the binary variables had the same value in all ten solutions. In addition, {\sc OneTree} generated a set of ten solutions of which only six solutions were unique. Among the six unique solutions, five of them had at most two variables with different values indicating very close similarity of the solutions. In constrast, {\sc DiversiTree}'s gave a solution set in which only 23 of the variables were the same in all ten solutions, all ten solutions were unique and at least nine variables had different values among the solutions, indicating the diversity of the solutions. This experiment provides further evidence supporting that the {\sc DiversiTree} approach provides a much more diverse set of train schedules from which a decision maker may select.

\section{Extensions}
\label{sec:6Extensions}

This study focused on investigating the diversity of solutions consisting of binary variables. While a majority of integer programming problems feature binary variables, there are some problems with pure integer variables (e.g., cutting stock optimization \citep{gilmore1961linear}) that could also benefit from a having a diverse set of near-optimal solutions. In addition, in some contexts, a user may like to consider diversity with respect to continuous variables. While many integer programming formulations feature continuous variables that depend on binary variables in a somewhat subordinate manner (such as facility location and lot sizing), for many other problems, this may not hold. We now discuss how {\sc DiversiTree} can be employed for problems involving variables beyond binary integer.

\subsection{Reformulations}
One way to consider diversity among general integer and continuous variables, is through reformulation.\\
\noindent\textbf{Binary Expansion of Pure Integer Variables.}
First, for the case of bounded pure integer variables, one approach is to replace each pure integer variable with its binary expansion. 
Consider a single nonnegative integer variable $x \in \mathbb{Z}$ with upper bound $u$. By introducing $M$ binary variables $b_0, \dots, b_{M-1}$ we can represent $x$ with its binary expansion:

\begin{equation}
\label{eqn:2Binarization1}
\begin{split}
     x & = \sum_{j=0}^{M-1} 2^j \cdot b_j ,\\
\end{split}
\end{equation}\\

\noindent which requires the smallest $M$ such that $u \leq \sum_{j=0}^{M-1}2^j$ so as to use as few binary variables as possible \citep{watters1967reduction}. One potential drawback of this method is that the number of binary variables needed to represent a single integer variable increases logarithmically in the upper bound. 

\noindent  \textbf{Binary Discretization of Continuous Variables.}
It is also possible to map a continuous variable to the binary space. Consider a single continuous variable $x \in \mathbb{R}$, and for simplicity of exposition let $x \in [0,1]$. To represent $x$ within accuracy $\epsilon = 10^{-p}$, where $p$ is some positive integer, then we need no more than $K = \left \lceil \frac{\log 10}{\log 2} p \right \rceil$ binary variables $z_k \in \{0,1 \}$ are needed, $k = 1, \dots, K$ so that $x \approx \sum_{k=1}^K 2^{-k}z_k$. As suggested previously, a potential drawback is that the number of binary variables needed to represent a single variable increases logarithmically with respect to $p$.

\subsection{Alternate Diversity Metrics}

The method presented in this paper is not limited to a specific metric for computing solution diversity. While the results in this paper are based on the $DBin$ metric, other metrics can be used equivalent to maximizing the average elementwise variances for the binary variables. Directly generalizing $DBin$, \cite{danna2009select} defined $DAll$, a metric that maximizes the variance of each variable and combines them by elementwise averaging:

\begin{equation}
\label{eqn:2Dall}
\begin{split}
     DAll(\solutionset) & = \frac{1}{|\solutionset|} \sum_{i=1}^{|\solutionset|} \frac{\sigma_i^2(\solutionset)}{\mathcal{R}},  \\
\end{split}
\end{equation}\\

\noindent where $\sigma_i^2(\solutionset) = \frac{1}{|\solutionset|} \sum_{j \in \solutionset} \left (x_i^{(j)} - \frac{1}{|\solutionset|} \sum_{k \in \solutionset} x_i^{(k)} \right )^2$ is the variance of variable $x_i$ in any set of solutions $S$. The scaling factor $\mathcal{R} = \max _{j \in \solutionset} x_i^{(j)} - \min_{j \in \solutionset} x_i^{(j)}$ is the range of values within the bounds for variable $x_i$.

However, as \cite{danna2009select} point out, a potential issue with using $DAll$ is that the variables in the MIP need to have known bounds to scale the variances to prevent a single variable having an outsized influence. As a mitigation strategy, \cite{danna2009select} propose the MIP owner provide importance weights for the different variables, which could then be a scaling factor for the calculated variances.

\section{Conclusion}
\label{sec:5conclusion}
We introduced a novel approach to generate diverse, near-optimal solution sets by emphasizing solution diversity in the node selection step of a branch-and-bound algorithm. Our results reveal that our modified node selection rule yields up to 190\% better diversity than known methods for generating diverse solutions. Our methods provide a fast way to generate diverse sets of near-optimal solutions, useful where there is utility in having a diverse set of options for decision making. We presented several methods for emphasizing diversity in node selection rules and optimal parameters that yield the most diverse set of solutions for the problems tested. We identified four groups of optimal parameters for different problems by clustering the parameter groups giving the best diversity ($DBin$) on the training set. Each group sets an optimal solution cutoff value and also sets how much diversity and depth to consider in determining the next solution added to the set of diverse solutions. When tested on new problems, our method (using the identified optimal parameters) ran in similar time as regular node selection rules and gave solution pools that were significantly more diverse.

A positive result of our study was that the four parameter setting groups identified during the parameter tuning on the training set performed well on a previously unseen set of problems in a test set. Even so, understanding the relationship between parameter settings and different classes of problems such that we can select the best parameter for any problem with a minimal amount of tuning would be beneficial in practice. A second area of interest which is also an extension of this work is diversifying solution pools on select variables (see \cite{voll2015optimum}). In cases where a cluster of variables represents an attribute that is desirable in a machine learning algorithm (such as fairness or intelligence), it may be useful and informative to generate solutions pools that are diverse on only that attribute or interpret and cluster the generated solutions based the attribute. Finally, we only reported results using $DBin$ as a diversity metric. It would be useful to understand how the new node selection method performs when using other metrics for computing diversity. 

\section*{Acknowledgments}
Medal gratefully acknowledges funding from the Army Research Office (grant W911NF-21-1-0079). However, the views expressed in this study do no represent those of the US Government, the US Department of Defense, or the US Army.\\

\bibliographystyle{plainnat}
\bibliography{biblography.bib}

\begin{thebibliography}{50}
\providecommand{\natexlab}[1]{#1}
\providecommand{\url}[1]{\texttt{#1}}
\expandafter\ifx\csname urlstyle\endcsname\relax
  \providecommand{\doi}[1]{doi: #1}\else
  \providecommand{\doi}{doi: \begingroup \urlstyle{rm}\Url}\fi

\bibitem[Achterberg et~al.(2008)Achterberg, Heinz, and Koch]{Achterberg.2008}
Tobias Achterberg, Stefan Heinz, and Thorsten Koch.
\newblock Counting solutions of integer programs using unrestricted subtree
  detection.
\newblock In \emph{International Conference on Integration of Artificial
  Intelligence (AI) and Operations Research (OR) Techniques in Constraint
  Programming}, Lecture Notes in Computer Science, pages 278--282, 2008.
\newblock ISBN 9783540681540.
\newblock \doi{10.1007/978-3-540-68155-7\_22}.

\bibitem[Baste et~al.(2022)Baste, Fellows, Jaffke, Masa{\v{r}}{\'\i}k,
  de~Oliveira~Oliveira, Philip, and Rosamond]{Baste.2019}
Julien Baste, Michael~R Fellows, Lars Jaffke, Tom{\'a}{\v{s}}
  Masa{\v{r}}{\'\i}k, Mateus de~Oliveira~Oliveira, Geevarghese Philip, and
  Frances~A Rosamond.
\newblock Diversity of solutions: An exploration through the lens of
  fixed-parameter tractability theory.
\newblock \emph{Artificial Intelligence}, 303:\penalty0 103644, 2022.

\bibitem[Bertsimas et~al.(2016)Bertsimas, King, and
  Mazumder]{bertsimas2016best}
Dimitris Bertsimas, Angela King, and Rahul Mazumder.
\newblock Best subset selection via a modern optimization lens.
\newblock \emph{The annals of statistics}, 44\penalty0 (2):\penalty0 813--852,
  2016.

\bibitem[Bixby et~al.(1998)Bixby, Ceria, McZeal, and
  Savelsbergh]{BixbyCeriaMcZealSavelsbergh1998}
R.~E. Bixby, S.~Ceria, C.~M. McZeal, and M.~W.~P Savelsbergh.
\newblock An updated mixed integer programming library: {MIPLIB} 3.0.
\newblock \emph{Optima}, 58:\penalty0 12--15, 1998.

\bibitem[Biyouki and Hwangbo(2021)]{biyouki2021blind}
Sajjad~Amrollahi Biyouki and Hoon Hwangbo.
\newblock Blind image deblurring based on kernel mixture.
\newblock \emph{arXiv preprint arXiv:2101.06241}, 2021.

\bibitem[Boehmer and Niedermeier(2021)]{boehmer2021broadening}
Niclas Boehmer and Rolf Niedermeier.
\newblock Broadening the research agenda for computational social choice:
  Multiple preference profiles and multiple solutions.
\newblock In \emph{Proceedings of the 20$^{th}$ International Conference on
  Autonomous Agents and MultiAgent Systems}, pages 1--5, 2021.

\bibitem[Cai et~al.(2021)Cai, He, Xie, Feng, and Ma]{Cai.2021}
Qi~Cai, Linwei He, Yimin Xie, Ruoqiang Feng, and Jiaming Ma.
\newblock {Simple and effective strategies to generate diverse designs for
  truss structures}.
\newblock \emph{Structures}, 32:\penalty0 268--278, 2021.
\newblock ISSN 2352-0124.
\newblock \doi{10.1016/j.istruc.2021.03.010}.

\bibitem[Cameron et~al.(2021)Cameron, Charmot, and Pulaj]{Cameron.2021}
Thomas~R Cameron, Sebastian Charmot, and Jonad Pulaj.
\newblock On the linear ordering problem and the rankability of data.
\newblock \emph{arXiv preprint arXiv:2104.05816}, 2021.

\bibitem[Ceria et~al.(1998)Ceria, Cordier, Marchand, and
  Wolsey]{ceria1998cutting}
Sebastian Ceria, Cecile Cordier, Hugues Marchand, and Laurence~A Wolsey.
\newblock Cutting planes for integer programs with general integer variables.
\newblock \emph{Mathematical programming}, 81\penalty0 (2):\penalty0 201--214,
  1998.

\bibitem[Church and Baez(2020)]{Church.2020}
Richard~L Church and Carlos~A Baez.
\newblock {Generating optimal and near-optimal solutions to facility location
  problems}.
\newblock \emph{Environment and Planning B: Urban Analytics and City Science},
  47\penalty0 (6):\penalty0 1014--1030, 2020.
\newblock ISSN 2399-8083.
\newblock \doi{10.1177/2399808320930241}.

\bibitem[Danna and Woodruff(2009)]{danna2009select}
Emilie Danna and David~L Woodruff.
\newblock How to select a small set of diverse solutions to mixed integer
  programming problems.
\newblock \emph{Operations Research Letters}, 37\penalty0 (4):\penalty0
  255--260, 2009.

\bibitem[Danna et~al.(2007)Danna, Fenelon, Gu, and
  Wunderling]{danna2007generating}
Emilie Danna, Mary Fenelon, Zonghao Gu, and Roland Wunderling.
\newblock Generating multiple solutions for mixed integer programming problems.
\newblock In \emph{International Conference on Integer Programming and
  Combinatorial Optimization}, pages 280--294. Springer, 2007.

\bibitem[Eckstein(1996)]{eckstein1996parallel}
Jonathan Eckstein.
\newblock Parallel branch-and-bound methods for mixed integer programming.
\newblock In \emph{Applications on Advanced Architecture Computers}, pages
  141--153. SIAM, 1996.

\bibitem[Eysenbach et~al.(2018)Eysenbach, Gupta, Ibarz, and
  Levine]{eysenbach2018diversity}
Benjamin Eysenbach, Abhishek Gupta, Julian Ibarz, and Sergey Levine.
\newblock Diversity is all you need: Learning skills without a reward function.
\newblock \emph{arXiv preprint arXiv:1802.06070}, 2018.

\bibitem[Gamrath et~al.(2020)Gamrath, Anderson, Bestuzheva, Chen, Eifler,
  Gasse, Gemander, Gleixner, Gottwald, Halbig, Hendel, Hojny, Koch, Le~Bodic,
  Maher, Matter, Miltenberger, M{\"u}hmer, M{\"u}ller, Pfetsch, Schl{\"o}sser,
  Serrano, Shinano, Tawfik, Vigerske, Wegscheider, Weninger, and
  Witzig]{GamrathEtal2020OO}
Gerald Gamrath, Daniel Anderson, Ksenia Bestuzheva, Wei-Kun Chen, Leon Eifler,
  Maxime Gasse, Patrick Gemander, Ambros Gleixner, Leona Gottwald, Katrin
  Halbig, Gregor Hendel, Christopher Hojny, Thorsten Koch, Pierre Le~Bodic,
  Stephen~J. Maher, Frederic Matter, Matthias Miltenberger, Erik M{\"u}hmer,
  Benjamin M{\"u}ller, Marc~E. Pfetsch, Franziska Schl{\"o}sser, Felipe
  Serrano, Yuji Shinano, Christine Tawfik, Stefan Vigerske, Fabian Wegscheider,
  Dieter Weninger, and Jakob Witzig.
\newblock {The SCIP Optimization Suite 7.0}.
\newblock Technical report, Optimization Online, March 2020.
\newblock URL
  \url{http://www.optimization-online.org/DB_HTML/2020/03/7705.html}.

\bibitem[Gilmore and Gomory(1961)]{gilmore1961linear}
Paul~C Gilmore and Ralph~E Gomory.
\newblock A linear programming approach to the cutting-stock problem.
\newblock \emph{Operations research}, 9\penalty0 (6):\penalty0 849--859, 1961.

\bibitem[Gleixner et~al.(2021)Gleixner, Hendel, Gamrath, Achterberg, Bastubbe,
  Berthold, Christophel, Jarck, Koch, Linderoth, L\"ubbecke, Mittelmann,
  Ozyurt, Ralphs, Salvagnin, and Shinano]{MIPLIB2017}
Ambros Gleixner, Gregor Hendel, Gerald Gamrath, Tobias Achterberg, Michael
  Bastubbe, Timo Berthold, Philipp~M. Christophel, Kati Jarck, Thorsten Koch,
  Jeff Linderoth, Marco L\"ubbecke, Hans~D. Mittelmann, Derya Ozyurt, Ted~K.
  Ralphs, Domenico Salvagnin, and Yuji Shinano.
\newblock {MIPLIB 2017: Data-Driven Compilation of the 6th Mixed-Integer
  Programming Library}.
\newblock \emph{Mathematical Programming Computation}, 2021.
\newblock \doi{10.1007/s12532-020-00194-3}.
\newblock URL \url{https://doi.org/10.1007/s12532-020-00194-3}.

\bibitem[Glover et~al.(1998)Glover, Kuo, and Dhir]{Glover.1998}
Fred Glover, Ching-Chung Kuo, and Krishna~S. Dhir.
\newblock {Heuristic algorithms for the maximum diversity problem}.
\newblock \emph{Journal of Information and Optimization Sciences}, 19\penalty0
  (1):\penalty0 109--132, 1998.
\newblock ISSN 0252-2667.
\newblock \doi{10.1080/02522667.1998.10699366}.

\bibitem[Glover et~al.(2000)Glover, L{\o}kketangen, and
  Woodruff]{glover2000scatter}
Fred Glover, Arne L{\o}kketangen, and David~L Woodruff.
\newblock Scatter search to generate diverse {MIP} solutions.
\newblock In \emph{Computing Tools for Modeling, Optimization and Simulation},
  pages 299--317. Springer, 2000.

\bibitem[Greistorfer et~al.(2008)Greistorfer, Løkketangen, Voß, and
  Woodruff]{Greistorfer.2008}
Peter Greistorfer, Arne Løkketangen, Stefan Voß, and David~L. Woodruff.
\newblock Experiments concerning sequential versus simultaneous maximization of
  objective function and distance.
\newblock \emph{Journal of Heuristics}, 14\penalty0 (6):\penalty0 613--625,
  2008.
\newblock ISSN 1381-1231.
\newblock \doi{10.1007/s10732-007-9053-z}.

\bibitem[He et~al.(2020)He, Cai, Zhao, and Xie]{He.2020}
Yunzhen He, Kun Cai, Zi-Long Zhao, and Yi~Min Xie.
\newblock {Stochastic approaches to generating diverse and competitive
  structural designs in topology optimization}.
\newblock \emph{Finite Elements in Analysis and Design}, 173:\penalty0 103399,
  2020.
\newblock ISSN 0168-874X.
\newblock \doi{10.1016/j.finel.2020.103399}.

\bibitem[Joseph et~al.(2015)Joseph, Dasgupta, Tuo, and
  Wu]{joseph2015sequential}
V~Roshan Joseph, Tirthankar Dasgupta, Rui Tuo, and CF~Jeff Wu.
\newblock Sequential exploration of complex surfaces using minimum energy
  designs.
\newblock \emph{Technometrics}, 57\penalty0 (1):\penalty0 64--74, 2015.

\bibitem[Koch et~al.(2011)Koch, Achterberg, Andersen, Bastert, Berthold, Bixby,
  Danna, Gamrath, Gleixner, Heinz, Lodi, Mittelmann, Ralphs, Salvagnin, Steffy,
  and Wolter]{KochEtAl2011}
Thorsten Koch, Tobias Achterberg, Erling Andersen, Oliver Bastert, Timo
  Berthold, Robert~E. Bixby, Emilie Danna, Gerald Gamrath, Ambros~M. Gleixner,
  Stefan Heinz, Andrea Lodi, Hans Mittelmann, Ted Ralphs, Domenico Salvagnin,
  Daniel~E. Steffy, and Kati Wolter.
\newblock {MIPLIB} 2010.
\newblock \emph{Mathematical Programming Computation}, 3\penalty0 (2):\penalty0
  103--163, 2011.
\newblock \doi{10.1007/s12532-011-0025-9}.
\newblock URL \url{http://mpc.zib.de/index.php/MPC/article/view/56/28}.

\bibitem[Kumar et~al.(2020)Kumar, Kumar, Levine, and Finn]{kumar2020one}
Saurabh Kumar, Aviral Kumar, Sergey Levine, and Chelsea Finn.
\newblock One solution is not all you need: Few-shot extrapolation via
  structured {M}ax{E}nt {RL}.
\newblock \emph{Advances in Neural Information Processing Systems}, 33, 2020.

\bibitem[Kuo et~al.(1993)Kuo, Glover, and Dhir]{kuo1993analyzing}
Ching-Chung Kuo, Fred Glover, and Krishna~S Dhir.
\newblock Analyzing and modeling the maximum diversity problem by zero-one
  programming.
\newblock \emph{Decision Sciences}, 24\penalty0 (6):\penalty0 1171--1185, 1993.

\bibitem[Lavine(2019)]{lavine2019whim}
Michael Lavine.
\newblock Whim: function approximation where it matters.
\newblock \emph{Communications in Statistics-Simulation and Computation}, pages
  1--31, 2019.

\bibitem[Lee et~al.(2000)Lee, Phalakornkule, Domach, and
  Grossmann]{lee2000recursive}
Sangbum Lee, Chan Phalakornkule, Michael~M Domach, and Ignacio~E Grossmann.
\newblock Recursive milp model for finding all the alternate optima in lp
  models for metabolic networks.
\newblock \emph{Computers \& Chemical Engineering}, 24\penalty0 (2-7):\penalty0
  711--716, 2000.

\bibitem[Liebchen and M{\"o}hring(2003)]{timtab1}
Christian Liebchen and Rolf~H M{\"o}hring.
\newblock Information on {MIPLIB}'s timetab-instances.
\newblock 2003.

\bibitem[Liebchen and Peeters(2002)]{liebchen2002cyclic}
Christian Liebchen and Leon Peeters.
\newblock On cyclic timetabling and cycles in graphs.
\newblock 2002.

\bibitem[Mouret and Clune(2015)]{mouret2015illuminating}
Jean-Baptiste Mouret and Jeff Clune.
\newblock Illuminating search spaces by mapping elites.
\newblock \emph{arXiv preprint arXiv:1504.04909}, 2015.

\bibitem[Pedregosa et~al.(2011)Pedregosa, Varoquaux, Gramfort, Michel, Thirion,
  Grisel, Blondel, Prettenhofer, Weiss, Dubourg, Vanderplas, Passos,
  Cournapeau, Brucher, Perrot, and Duchesnay]{scikit-learn}
F.~Pedregosa, G.~Varoquaux, A.~Gramfort, V.~Michel, B.~Thirion, O.~Grisel,
  M.~Blondel, P.~Prettenhofer, R.~Weiss, V.~Dubourg, J.~Vanderplas, A.~Passos,
  D.~Cournapeau, M.~Brucher, M.~Perrot, and E.~Duchesnay.
\newblock Scikit-learn: Machine learning in {P}ython.
\newblock \emph{Journal of Machine Learning Research}, 12:\penalty0 2825--2830,
  2011.

\bibitem[Petit and Trapp(2015)]{petit2015finding}
Thierry Petit and Andrew~C Trapp.
\newblock Finding diverse solutions of high quality to constraint optimization
  problems.
\newblock In \emph{Twenty-Fourth International Joint Conference on Artificial
  Intelligence}, 2015.

\bibitem[Petit and Trapp(2019)]{petit2019enriching}
Thierry Petit and Andrew~C Trapp.
\newblock Enriching solutions to combinatorial problems via solution
  engineering.
\newblock \emph{INFORMS Journal on Computing}, 31\penalty0 (3):\penalty0
  429--444, 2019.

\bibitem[Rodríguez-Mier et~al.(2021)Rodríguez-Mier, Poupin, Blasio, Cam, and
  Jourdan]{Rodriguez-Mier.2021}
Pablo Rodríguez-Mier, Nathalie Poupin, Carlo~de Blasio, Laurent~Le Cam, and
  Fabien Jourdan.
\newblock {DEXOM: Diversity-based enumeration of optimal context-specific
  metabolic networks}.
\newblock \emph{PLOS Computational Biology}, 17\penalty0 (2):\penalty0
  e1008730, 2021.
\newblock ISSN 1553-734X.
\newblock \doi{10.1371/journal.pcbi.1008730}.

\bibitem[Rudin et~al.(2022)Rudin, Chen, Chen, Huang, Semenova, and
  Zhong]{rudin2022interpretable}
Cynthia Rudin, Chaofan Chen, Zhi Chen, Haiyang Huang, Lesia Semenova, and Chudi
  Zhong.
\newblock Interpretable machine learning: Fundamental principles and 10 grand
  challenges.
\newblock \emph{Statistics Surveys}, 16:\penalty0 1--85, 2022.

\bibitem[Schittekat and S{\"o}rensen(2009)]{schittekat2009or}
Patrick Schittekat and Kenneth S{\"o}rensen.
\newblock {OR} practice—supporting {3PL} decisions in the automotive industry
  by generating diverse solutions to a large-scale location-routing problem.
\newblock \emph{Operations Research}, 57\penalty0 (5):\penalty0 1058--1067,
  2009.

\bibitem[Schwind et~al.(2020)]{schwind2020representative}
Nicolas Schwind et~al.
\newblock Representative solutions for bi-objective optimisation.
\newblock In \emph{Proceedings of the AAAI Conference on Artificial
  Intelligence}, volume~34, pages 1436--1443, 2020.

\bibitem[Semenova et~al.(2019)Semenova, Rudin, and Parr]{semenova2019study}
Lesia Semenova, Cynthia Rudin, and Ronald Parr.
\newblock A study in {R}ashomon curves and volumes: A new perspective on
  generalization and model simplicity in machine learning.
\newblock \emph{arXiv preprint arXiv:1908.01755}, 2019.

\bibitem[Serra(2020)]{Serra.2020}
Thiago Serra.
\newblock Enumerative branching with less repetition.
\newblock In \emph{International Conference on Integration of Constraint
  Programming, Artificial Intelligence, and Operations Research}, Lecture Notes
  in Computer Science, pages 399--416, 2020.
\newblock ISBN 9783030589417.
\newblock \doi{10.1007/978-3-030-58942-4\_26}.

\bibitem[Serra and Hooker(2020)]{10.1007/s10107-019-01390-3}
Thiago Serra and J.~N. Hooker.
\newblock {Compact representation of near-optimal integer programming
  solutions}.
\newblock \emph{Mathematical Programming}, 182\penalty0 (1-2):\penalty0
  199--232, 2020.
\newblock ISSN 0025-5610.
\newblock \doi{10.1007/s10107-019-01390-3}.

\bibitem[Sharifnia et~al.(2021)Sharifnia, Biyouki, Sawhney, and
  Hwangbo]{sharifnia2021robust}
Seyed Mohammad~Ebrahim Sharifnia, Sajjad~Amrollahi Biyouki, Rupy Sawhney, and
  Hoon Hwangbo.
\newblock Robust simulation optimization for supply chain problem under
  uncertainty via neural network metamodeling.
\newblock \emph{Computers \& Industrial Engineering}, 162:\penalty0 107693,
  2021.

\bibitem[Trapp and Konrad(2015)]{trapp2015finding}
Andrew~C Trapp and Renata~A Konrad.
\newblock Finding diverse optima and near-optima to binary integer programs.
\newblock \emph{IIE Transactions}, 47\penalty0 (11):\penalty0 1300--1312, 2015.

\bibitem[Tsoupidi et~al.(2020)Tsoupidi, Lozano, and Baudry]{Tsoupidi.2020}
Rodothea~Myrsini Tsoupidi, Roberto~Castañeda Lozano, and Benoit Baudry.
\newblock {Principles and Practice of Constraint Programming, 26$^{th}$
  International Conference, CP 2020, Louvain-la-Neuve, Belgium, September
  7–11, 2020, Proceedings}.
\newblock \emph{Lecture Notes in Computer Science}, pages 791--808, 2020.
\newblock ISSN 0302-9743.
\newblock \doi{10.1007/978-3-030-58475-7\_46}.

\bibitem[Van~Hentenryck et~al.(2009)Van~Hentenryck, Coffrin, and
  Gutkovich]{van2009constraint}
Pascal Van~Hentenryck, Carleton Coffrin, and Boris Gutkovich.
\newblock Constraint-based local search for the automatic generation of
  architectural tests.
\newblock In \emph{International Conference on Principles and Practice of
  Constraint Programming}, pages 787--801. Springer, 2009.

\bibitem[Voll et~al.(2015)Voll, Jennings, Hennen, Shah, and
  Bardow]{voll2015optimum}
Philip Voll, Mark Jennings, Maike Hennen, Nilay Shah, and Andr{\'e} Bardow.
\newblock The optimum is not enough: A near-optimal solution paradigm for
  energy systems synthesis.
\newblock \emph{Energy}, 82:\penalty0 446--456, 2015.

\bibitem[Wang et~al.(2018)Wang, Zhou, Zhou, Xu, and Niu]{wang2018diverse}
Bo~Wang, Yan Zhou, Yiming Zhou, Shengli Xu, and Bin Niu.
\newblock Diverse competitive design for topology optimization.
\newblock \emph{Structural and Multidisciplinary Optimization}, 57\penalty0
  (2):\penalty0 891--902, 2018.

\bibitem[Watters(1967)]{watters1967reduction}
Lawrence~J Watters.
\newblock Reduction of integer polynomial programming problems to zero-one
  linear programming problems.
\newblock \emph{Operations Research}, 15\penalty0 (6):\penalty0 1171--1174,
  1967.

\bibitem[Zahavy et~al.(2021)Zahavy, O'Donoghue, Barreto, Mnih, Flennerhag, and
  Singh]{zahavy2021discovering}
Tom Zahavy, Brendan O'Donoghue, Andre Barreto, Volodymyr Mnih, Sebastian
  Flennerhag, and Satinder Singh.
\newblock Discovering diverse nearly optimal policies with successor features.
\newblock \emph{arXiv preprint arXiv:2106.00669}, 2021.

\bibitem[Zaslavsky and Singh(2006)]{Zaslavsky.2006}
Elena Zaslavsky and Mona Singh.
\newblock {A combinatorial optimization approach for diverse motif finding
  applications}.
\newblock \emph{Algorithms for Molecular Biology}, 1\penalty0 (1):\penalty0
  1--13, 2006.
\newblock \doi{10.1186/1748-7188-1-13}.

\bibitem[Zhou et~al.(2016)Zhou, Haftka, and Cheng]{Zhou.2016}
Yiming Zhou, Raphael~T. Haftka, and Gengdong Cheng.
\newblock Balancing diversity and performance in global optimization.
\newblock \emph{Structural and Multidisciplinary Optimization}, 54\penalty0
  (4):\penalty0 1093--1105, 2016.
\newblock ISSN 1615-147X.
\newblock \doi{10.1007/s00158-016-1434-1}.

\end{thebibliography}

\clearpage

\begin{appendices}
\section{Problems in the training set}
\label{Sec:trainingSet}
Table \ref{fig:problemsSolved2} captures the problems we solved in the training set. They are randomly selected problems from MIPLIB \citep{MIPLIB2017, KochEtAl2011, BixbyCeriaMcZealSavelsbergh1998}. We did remove instances that did not complete computation of the objective value within 30 minutes. In total, 27 problem instances are used in this training set. We capture the characteristics of the problem below.

{\small
\begin{xltabular}[H]{\linewidth}{@{ }X @{ }X @{ }X @{ }X @{ }X @{ }X}
\captionsetup{margin=0.5cm}
    \caption{\textbf{The problem instances used for training in \S \ref{sec:diversity-of-common-node-selection} and \S \ref{sec:parameter-optimization} and their characteristics are shown in this table.}}
    \label{fig:problemsSolved2}\\
   \hline
  Problem Instance  & Total \newline variables & Binary \newline variables &  General \newline integer \newline variables  & Continuous variables  & Solutions within 1\% of the optimum \\  
  \hline
  \endfirsthead
  \multicolumn{6}{ l }{Table: \ref{fig:problemsSolved2} Continued} \\
  \hline
  \endhead
  
   air03  & 10757 & 10757 & 0 & 0 &938 \\
    bell5  & 104 & 30 & 28 & 46 & $>$10,000 \\
    dcmulti  & 548 & 75 & 0 & 473 & $>$10,000 \\
    fiber  & 1298 & 1254 & 0 & 44 & $>$279 \\
    fixnet6  & 878 & 378 & 0 & 500 & $>$10,000 \\
    gen  & 870  & 144 & 6 & 720 & $>$10,000 \\
   gesa3  & 1152  & 216 & 168 & 768 & $>$10,000 \\
   gt2  & 188  & 24 & 164 & 0 & $>$10,000 \\
   khb05250  & 1350  & 24 & 0 & 1326 & 28 \\
   l152lav  & 1989  & 1989 & 0 & 0 & $>$10,000 \\
   misc03  & 160  & 159 & 0 & 1 & 24 \\
   misc06  & 1808  & 112 & 0 & 1696 & $>$10,000 \\
   mod008  & 319  & 319 & 0 & 0 & 68 \\
   mod010  & 2655  & 2655 & 0 & 0 & $>$10,000 \\
   p0033  & 33  & 33 & 0 & 0 & 15 \\
   p0201  & 201  & 201 & 0 & 0 & 44 \\
   p0548  & 548  & 548 & 0 & 0 & $>$10,000 \\
   pp08a  & 240  & 64 & 0 & 176 & 64 \\
   pp08aCUTS  & 240  & 64 & 0 & 176 & 64 \\
   $qnet1_o$  & 1541  & 1288 & 129 & 124 & $>$10,000 \\
   qnet1  & 1541  & 1288 & 129 & 124 & $>$10,000 \\
   rgn  & 180  & 100 & 0 & 90 & $>$720 \\
   set1ch  & 712  & 240 & 0 & 472 & $>$10,000 \\
   stein27  & 27  & 0 & 0 & 0 & 2106 \\
   stein45  & 45  & 0 & 0 & 0 & 70 \\
  \hline
   vpm1  & 378  & 168 & 0 & 210 & $>$10,000 \\
   vpm2  & 378  & 168 & 0 & 210 & 33 \\
  \hline
 \end{xltabular}
}

\section{Problems in the testing set}
\label{Sec:testingSet}
Table \ref{fig:problemsSolved3} captures the problems we solved in the testing set. They are randomly selected problems from MIPLIB \citep{MIPLIB2017}. We did remove instances that did not complete computation of the objective value within 30 minutes. In total, 9 problem instances are used in this training set. We capture the characteristics of the problem below.

{\small
\begin{xltabular}[H]{\linewidth}{@{ }X @{ }X @{ }X @{ }X @{ }X @{ }X}
\captionsetup{margin=0.5cm}
    \caption{\textbf{The problem instances used for testing in \S \ref{sec:diversity-of-common-node-selection} and \S \ref{sec:parameter-optimization} and their characteristics are shown in this table.}}
    \label{fig:problemsSolved3}\\
   \hline
  Problem Instance  & Total \newline variables & Binary \newline variables &  General \newline integer \newline variables  & Continuous variables  & Solutions within 1\% of the optimum \\  
  \hline
  \endfirsthead
  \multicolumn{6}{ l }{Table: \ref{fig:problemsSolved3} Continued} \\
  \hline
  \endhead
  
   23588  & 368 & 231 & 0 & 137 & 82 \\
    bppc8-02  & 232 & 229 & 1 & 2 & $>$10,000 \\
    exp-1-500-5-5  & 990 & 250 & 0 & 740 & $>$1,338 \\
    mtest4ma  & 1950 & 975 & 0 & 975 & $>$10,000 \\
    neos-1425699  & 105 & 5 & 80 & 20 & $>$10,000 \\
    neos17  & 535  & 300 & 0 & 235 & $>$10,000 \\
   nexp-50-20-1-1  & 490  & 245 & 0 & 245 & $>$10,000 \\
   sp150x300d  & 600  & 300 & 0 & 300 & $>$9,455 \\
  \hline
 \end{xltabular}
 }
 
 \section{Parameter groups for diversity-emphasizing rules}
\label{Sec:paramGroups}

\begin{figure}[H]
    \centering
    \includegraphics[width=\linewidth]{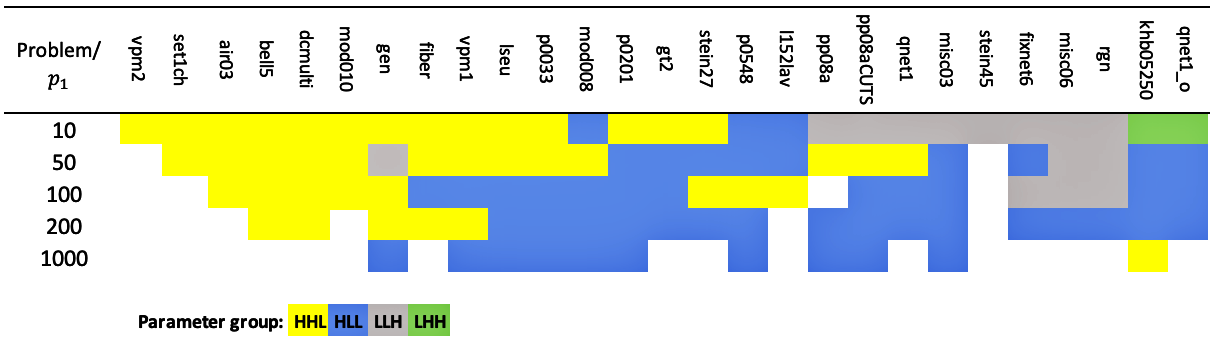}
    \captionsetup{margin=2cm,format=plain, font=small, singlelinecheck=false}
    \caption{The four parameter groups and the problem instances in each group as the number of requested solutions increases from 10 to 1,000.}
    \label{fig:parameterGroups}
\end{figure}

Figure \ref{fig:parameterGroups} shows the clustering of the problem instances into the four groups. As the figure shows, these four groups consolidate to three (HHL, HLL, and LLH) when the number of requested solutions reaches 50 and to two (HHL and HLL) for 200 requested solutions or more. The consolidation to groups HHL and HLL indicate that emphasizing diversity after generating a seed set of solutions results in a higher overall diversity of all generated solutions. 

One might suspect that two instances of the same problem (e.g., stein27 and stein45) would typically have the same parameter group. However, surprisingly we found that similar problems like stein27/stein45 and qnet1/qnet1\_0 did not always belong to the same group. At $p_1 \geq 50$, some problems took too long to complete all test cases in the grid search and thus were excluded from the parameter groups shown in Figure \ref{fig:parameterGroups}.

\end{appendices}

\end{document}